\documentstyle[12pt,aaspp4]{article}

\begin{document}

\title{A Far Ultraviolet Analysis of the Stellar Populations in
Six Elliptical and S0 Galaxies}

\author{Thomas M. Brown}

\affil{Department of Physics and Astronomy, Johns Hopkins University\\ 
Charles and 34th Streets, Baltimore, MD 21218}

\author{Henry C. Ferguson}
\affil{Space Telescope Science Institute\\
3700 San Martin Drive, Baltimore, MD 21218}

\author{Arthur F. Davidsen}
\affil{Department of Physics and Astronomy, Johns Hopkins University\\ 
Charles and 34th Streets, Baltimore, MD 21218}

\and

\author{Ben Dorman}
\affil{Laboratory for Astronomy \& Solar Physics, NASA/GSFC\\
Greenbelt, MD 20771}

\begin{center}
To appear in \it The Astrophysical Journal \rm
\end{center}

\begin{abstract}

We have analyzed the far-ultraviolet (FUV) spectra of six elliptical and
S0 galaxies in order to characterize the stellar population that produces the 
ultraviolet flux in these galaxies.  The spectra were obtained using the 
Hopkins Ultraviolet Telescope (HUT) during the Astro-2 mission aboard the 
Space Shuttle \it Endeavour \rm 
in March 1995, and cover the spectral range from 820 
to 1840~\AA\ with a resolution of 3~\AA.  These data, together with the 
spectra of two galaxies observed with HUT on the Astro-1 mission, represent 
the only FUV spectra of early type galaxies that extend to the Lyman limit at 
912~\AA\ and therefore include the ``turnover'' in the spectral energy 
distribution below Lyman alpha.
 
Using an extensive new grid of LTE and non-LTE synthetic spectra which match 
the HUT resolution and cover the relevant parameter space of temperature and 
gravity, we have constructed synthetic 
spectral energy distributions by integrating over various predicted stellar 
evolutionary tracks for horizontal branch stars and their progeny. 
When the computed models are compared with the HUT data, we find that 
models with supersolar metal abundances and helium 
best reproduce the flux across the entire HUT 
wavelength range, while those with subsolar $Z$ \& $Y$ fit less well, partly
because of a significant flux deficit shortward of 970~\AA\ in the models.  
High $Z$ models are preferred because the contribution
from the later, hotter, post-HB evolutionary stages makes
up a higher fraction of the sub-Lyman~$\alpha$ flux in these tracks.
We find that AGB-Manqu$\acute{\rm e}$ 
evolution is required in all of the fits to the HUT spectra,
suggesting that all of the galaxies have some
subdwarf B star population.  At any $Z$ \& $Y$, 
the model spectra that best match the HUT flux are dominated by
stars evolving from a narrow range of envelope mass on the blue end
of the horizontal branch.
 
The Astro-1 and Astro-2 data are also the first with the resolution and 
signal-to-noise needed to detect and measure absorption lines in the FUV 
spectra of elliptical galaxies, allowing a direct estimate of the abundances
in the atmospheres of the stars that produce the UV flux.  We find that most 
absorption features in the spectra are consistent with 
$Z = 0.1$~$Z_{\sun}$, significantly lower than the abundances implied by the 
best-fitting spectral energy distributions.  However, given the strong 
observational and theoretical evidence for diffusion processes in the 
atmospheres of evolved stars, the observed atmospheric abundances may not
reflect the interior 
abundances in the population producing the ultraviolet flux 
in elliptical galaxies.

\end{abstract}


\section{INTRODUCTION}

An understanding of the stellar populations in elliptical galaxies 
will shed light upon conflicting theories in cosmology, galactic
evolution, and stellar evolution.    
However, 25 years after the initiation of space-based observations and
the consequent extension of data into the ultraviolet (UV), there are still
considerable uncertainties regarding the chemical composition 
in elliptical galaxies and the evolution of their stellar populations.

The spectra of elliptical galaxies and spiral galaxy bulges exhibit
a strong upturn shortward of 2700~\AA, dubbed the ``UV upturn.''
Characterized by the ($m_{1550}-V$) color, the
UV upturn shows strong variation (ranging from 2.05--4.50 mag) in 
nearby quiescent early-type galaxies (Burstein et al.\ 1988\markcite{B88}).
The strength of the UV upturn
is positively correlated with the strength of Mg$_2$ line absorption
in the $V$ band, in the sense that the ($m_{1550}-V$) color is bluer at
higher line strengths, opposite to the behavior of optical color indices
(Burstein et al.\ 1988\markcite{B88}).  
Opposing theories have been devised to explain this correlation.  In one camp,
Lee (1994\markcite{L94}) and Park \& Lee (1997\markcite{PL97})
suggest that the UV flux originates in the
low metallicity tail of an evolved stellar population with
a wide metallicity distribution.  In the other camp, several groups
(Bressan, Chiosi, \& Fagotto 1994\markcite{BCF94};
Greggio \& Renzini 1990\markcite{GR90};
Horch, Demarque, \& Pinsonneault 1992\markcite{HDP92}) propose that metal-rich
horizontal branch (HB) stars and their progeny are responsible for the UV
flux.  These different metallicity scenarios in turn argue for different
ages for the stellar populations in these galaxies.  Ages exceeding those
of Galactic globular clusters are required in the Park \& Lee
(1997\markcite{PL97}) model, while ages as low as 8 Gyr are allowed in 
the Bressan et al.\ (1994\markcite{BCF94}) model.

In these two scenarios,
the EHB stars are drawn from either 
tail of the metallicity distribution.  However,
it is also possible, indeed perhaps more likely, that the EHB stars
arise from progenitors near the peak of the metallicity distribution
(cf.\ Dorman, O'Connell, \& Rood 1995\markcite{DOR95}),
but represent a relatively rare occurrence.  
Observations of a bimodal
HB within the metal-rich open cluster NGC~6791 suggest that this might be the 
case (Liebert, Saffer, \& Green 1994\markcite{LSG94}).  If the same 
mechanism is at work in both the elliptical galaxies and in NGC~6791, 
the correlation of ($m_{1550}-V$) with the global metallicity of the galaxy
might indicate that this rare path of stellar evolution becomes less so
at high metallicities. 

\subsection{Horizontal Branch Morphology}

How can two diametrically opposed interpretations arise from the same
observational data?  The answer lies in the parameters that govern HB 
morphology,
such as metallicity.  In general, for stars of solar metallicity or less
([Fe/H] $< 0$), 
higher metallicities yield redder HBs and lower metallicities yield bluer HBs,
i.e. bluer HBs have more stars on the blue side of
the RR Lyrae instability strip, and redder HBs have more
stars on the red side.   Metallicity
is the implied ``first parameter'' of HB morphology. 
In his spectrophotometry of 13 Milky Way globular clusters,
Gregg (1994\markcite{G94}) gives an excellent observational example
of the classical effect of metallicity on HB morphology; 
Lee, Demarque, \& Zinn (1994\markcite{LDZ94})
provide theoretical examples.  
However, it has been known for decades that other parameters affect
HB morphology (cf.\ Lee et al.\ 1994\markcite{LDZ94} and references therein).
In this way, the metallicity debate in elliptical galaxies
is tied to the ``second parameter'' debate that has been raging in
the study of globular clusters.

Helium abundance also plays an important role in the stellar structure of
zero-age HB (ZAHB) stars (Dorman 1992\markcite{D92}).  These stars
are characterized by a helium-rich core and a hydrogen-rich 
envelope.  At the core/envelope boundary, the density $\rho$ and mean molecular
weight $\mu$ change such that $\rho/\mu$ is continuous.  As the helium 
abundance ($Y$) in the envelope increases, the $\mu$ gradient across the 
core/envelope boundary decreases, which in turn decreases the temperature 
gradient across this boundary.  Shallower temperature gradients produce higher
shell temperatures, and the nuclear reactions in the hydrogen-burning
shell are highly 
sensitive to the temperature.  At the red end of the ZAHB, the shell 
luminosity in these massive envelopes makes a significant contribution to the 
total luminosity, and so an increase in temperature results in a significant 
increase in the total luminosity, but does not change the envelope structure
significantly.  At the blue end of the ZAHB, the shell luminosity in these 
less massive envelopes makes a smaller contribution to the total luminosity, 
and so the total luminosity does not change significantly with increasing
temperature, but the envelope structure does.  Thus, an increase in $Y$ on the 
blue end of the ZAHB significantly increases T$_{\rm eff}$ and slightly 
increases the luminosity, while an increase in $Y$ on the red end of the ZAHB 
significantly increases the luminosity and slightly increases T$_{\rm eff}$.  
Across the entire ZAHB, an increase in $Y$ increases the rate of nuclear 
reactions, resulting in more rapid consumption of the envelope.

Lee et al.\ (1994\markcite{LDZ94}) argue that age is the second parameter
driving HB morphology (see also Sarajedini, Lee, \& Lee
1995\markcite{SLL95}).  They also investigate
the role of He abundance, 
CNO/Fe abundance ratio, and core rotation,
but rule out these latter candidates for the second parameter because of 
conflicts with observations.  Lee et al.\ (1994\markcite{LDZ94}) find
that age variations of several Gyr can produce dramatic differences
in HB morphology, in the sense that young HBs (age~=~11--12 Gyr)
tend to be red, while older HBs (age~=~15--16 Gyr) tend to be blue.
This is because the mass of a red giant at the helium flash decreases
as the age of a stellar population increases, which in turn 
decreases the mean mass of the HB stars (Lee et al.\ 1994\markcite{LDZ94} and 
references therein).  Although this scenario is true in any given system,
it must be stressed that when comparing different populations, the
total mass loss on the red giant branch (RGB) may not have the same mean value
and dispersion, and thus differences in HB morphology cannot 
necessarily be attributed
to differences in age alone.  When more than one parameter drives the HB
mass distribution, it may be difficult to separate age from these other 
effects.

For an HB with a single metallicity, the envelope mass ($M_{env}$)
determines the color location of stars on the HB.
According to theory, stars with
lower envelope mass are bluer (hotter) and stars with 
higher envelope mass are redder (cooler).  Stars evolving off the HB
at core helium exhaustion
will then follow one of three different evolutionary scenarios, again
depending upon the envelope mass.  Dorman, Rood, \& O'Connell 
(1993\markcite{DRO93}, hereafter DRO93) explain these evolutionary scenarios 
in detail; we will briefly summarize them here.

The first scenario occurs for an HB star with 
$M_{env}\gtrsim 0.07$~$M_{\sun}$ (in the case of solar $Z$ and $Y$).
Such a star evolves along the asymptotic 
giant branch (AGB), 
and after reaching the thermally pulsing (TP) stage, much of
the envelope is removed and the star no longer sustains a deep convective
exterior.  At this point it evolves rapidly to higher temperatures at fixed 
luminosity, and then fades along the white dwarf (WD) cooling track.
An example of a classical
post-asymptotic giant branch (PAGB) evolutionary track 
(Vassiliadis \& Wood 1994\markcite{VW94}) is given in
Fig.~1, along with the integrated far-ultraviolet (FUV) 
spectrum of a population of stars entering that track at the rate of
one star per year.  An AGB path that might precede the PAGB evolution
is also shown (DRO93\markcite{DRO93}), but this cool, red portion
of the evolution (from HB to TP) makes a negligible contribution 
to the integrated FUV luminosity in the HUT wavelength range.

The second scenario occurs
if an HB star has enough envelope mass to reach the AGB, but not enough 
to maintain a convective exterior all the way to the TP stage
(0.03~$M_{\sun}$~$\lesssim M_{env} \lesssim$~0.07~$M_{\sun}$ for
$Y =$~$Y_{\sun}$ and $Z=$~$Z_{\sun}$).
These stars, dubbed post-early AGB (PEAGB; Brocato et 
al.\ 1990\markcite{BT90}), 
evolve away from the AGB before reaching the TP stage, moving
to higher temperatures at a constant luminosity which is less than that
in the PAGB scenario.  Because of the lower luminosity, 
and because models with lower envelope masses leave the AGB with
more remaining hydrogen fuel, the PEAGB stars
have longer lifetimes.  Thus they emit more UV light over those lifetimes, as
compared to the stars following the PAGB tracks.  Fig.~1 shows an example
of PEAGB evolution (DRO93\markcite{DRO93}) and the
corresponding FUV luminosity.  This particular PEAGB track exhibits
helium shell flash TP ``loops'' which take place
on the AGB if the envelope is more than a few hundredths of a solar mass.
Although they are dramatic in the HR diagram, 
these short-lived features do not significantly alter the integrated spectrum.

The third scenario occurs for 
those HB stars with such small envelope masses 
($M_{env} \lesssim$~0.03~$M_{\sun}$ for
$Y =$~$Y_{\sun}$ and $Z=$~$Z_{\sun}$) that they are unable ever to
 develop deep outer convection zones, and thus are unable to
reach the AGB at all.  
In the shell burning phase they are referred to as
AGB-Manqu$\acute{\rm e}$ (AGBM) stars, and they spend $\sim 10^7$~yr at
relatively high luminosity ($\sim 100$~$L_{\sun}$) and high 
temperature (T$_{\rm eff} > 30,000$~K), after which they evolve directly
to the WD cooling track.
The amount of FUV light emitted from AGBM stars is
much higher than that emitted from PAGB or PEAGB stars, as shown in Fig.~1,
again due to their longer evolutionary time scales.  
Following the nomenclature of DRO93\markcite{DRO93},
we will use the term extreme HB (EHB) for those HB stars that evolve 
along PEAGB or AGBM evolutionary paths.

The different evolutionary paths a star may follow after leaving the
HB are affected by the metallicity and helium abundance of the star.
By itself, the envelope opacity at high $Z$ has a tendency to ``damp down''
the hydrogen burning shell, so that high-metallicity stars are fainter
(Dorman 1992\markcite{D92}).  
It is usually supposed that enhanced metal
abundance is seen in stellar populations accompanied with increases in
the helium fraction, according to a simple linear formula: $\Delta
Y/\Delta Z =$~constant.  Values for this constant are, however,
determined from observations close to $Z = 0$ rather than $Z =
Z_{\sun}$, since what is often sought is the primordial helium abundance
for cosmological studies. There is little information about the helium
abundance of metal-rich populations and the subject is quite
controversial (see Dorman et al.\ 1995\markcite{DOR95} and references 
therein; for a recent study, see Minniti 1995\markcite{M95}).
For high He envelope abundance 
($Y \gtrsim 0.40$), the envelope is consumed at a higher rate,
because the hydrogen shell burns at higher temperatures (Dorman 
1992\markcite{D92}; Sweigart \& Gross 1976\markcite{SG76}).
The result is that more stars, with a greater range of envelope mass on 
the HB, will follow the PEAGB and AGBM paths (Fagotto
et al.\ 1994c\markcite{FBBC94c}; DRO93\markcite{DRO93}). 
It must be stressed that
it is the high {\it absolute} value of the helium abundance that causes
this effect (and not the increase in $Z$).
 
High metallicity might affect
the mass loss processes on the RGB.  Compared
to metal-poor RGB stars, metal-rich
RGB stars have lower masses and larger radii at the same luminosity,
and therefore lower surface gravities, which may in turn increase
the rate of mass loss on the RGB (Horch et al.\ 1992\markcite{HDP92} and
references therein).  With greater mass loss, more stars will arrive
on the EHB.  We note that although the increase of mass loss with increasing
metallicity is certainly plausible, the effect remains
unsubstantiated by theory or observations.

Once we understand the different physical parameters driving HB
morphology, we may see how two different 
interpretations can arise for the correlation between Mg$_2$ strength
and UV upturn strength.  
Arguing for the case of high metallicity, 
Bressan et al.\ (1994\markcite{BCF94}) say that more massive galaxies
have higher mean and \it maximum \rm metallicity, and thus
--- assuming $Y$ also becomes high for high $Z$ --- 
more stars have a high enough metallicity to follow the evolutionary
paths of the PEAGB and AGBM stars.  Taking the opposing view, 
Lee (1994\markcite{L94}) argues that more massive galaxies have
higher mean metallicity and are \it older \rm than less massive galaxies,
and thus whatever metal-poor fraction there is in
the stellar population will have bluer HBs.  
Measurements of FUV line strengths have been proposed as sensitive tests
of the models by both camps (see also Yi et al.\ 1995\markcite{Y95}).
In the Park \& Lee (1997\markcite{PL97}) model, one might expect to
observe weak \ion{C}{4} and \ion{Si}{4} features anticorrelated with
($m_{1550}-V$).  In the metal-rich scenario, one might expect strong
lines and a positive correlation with ($m_{1550}-V$).

The different metallicity scenarios have important implications for
the ages of elliptical galaxies and in turn the age of the Universe.
Bressan et al.\ (1994\markcite{BDF94})
demonstrated that in the high-$Z$ -- high-$Y$ scheme,
the onset of UV excess serves as an age probe,
since it is the most rapidly evolving feature in the spectrum
of an old elliptical galaxy.  In their example, Bressan et 
al.\ (1994\markcite{BCF94}) show that the onset of UV excess occurs
in the galaxy rest frame at an age of $7.6\times10^9$~yr.
In the low-metallicity scheme, Park \& Lee (1997\markcite{PL97}) claim
that giant elliptical galaxies must be at least 3 Gyr older than
our own Galaxy, implying the lower limit to the age of the 
Universe must be 19 Gyr.  
Both models assume a Gaussian mass dispersion and either a
Reimers (1975\markcite{R75}) mass loss law or fixed mass loss. 
The ages and metallicities
at which the UV upturn begins to appear are unfortunately quite
sensitive to these assumptions and to the parameters. The
production of hot HB stars in the globular clusters and
in the Galactic field is not easily described by a simple
function, as we do not know how to characterize mass
loss in stars.   The UV upturn therefore has little
value as an age indicator until the physics driving the phenomenon 
is understood.

In contrast, Dorman et al.\ (1995\markcite{DOR95}) 
argue that the size of the population of EHB stars --- i.e.,
the fraction of all HB stars in the galaxy that are UV-bright --- 
is the only sure deduction that can be made from the FUV radiation.
They estimate, by using broadband colors derived from simple models of 
all evolutionary phases, that the blue FUV sources must have
EHB fractions reaching $\sim 20\, \%$ of the HB population, a result only
weakly dependent on the $Z-$abundance of the models. They argue
that such a fraction is not a ``trace population.'' Instead,
it is more plausible that the UV flux is produced by a significant
minority of the dominant population, comprised of stars which have
undergone higher mass loss than their red HB counterparts. 
The reason for the UV-Mg$_2$ correlation remains unknown, but may
be explained by an increase in the likelihood of high mass loss in
more metal-rich (or, at least, $\alpha-$ enhanced) stellar populations.
In this picture there is no way to relate
the FUV radiation to 
galaxy age, until mass loss in HB stars is better understood.
In addition, the UV upturn is
correlated with the velocity dispersion ($\sigma_v$), 
and although the correlation
between ($m_{1550}-V$) and Mg$_2$ index is tighter than that between 
($m_{1550}-V$) and $\sigma_v$
(Burstein et al.\ 1988\markcite{B88}),
it is not yet confirmed that the underlying correlation
(i.e. causal relationship) is with abundance.

Because of the great uncertainties in the physical mechanisms responsible
for the production of EHB stars, we have chosen in this paper to look
at the constraints provided by the UV spectra themselves, with only
very broad constraints on the overall stellar populations in the galaxies.
Our conclusions are thus independent of any particular model of elliptical
galaxy evolution.  The limitation of this approach is that without a model
there is no straightforward way to specify the distribution of stars on
the zero-age HB.  This would be a serious problem if there were a wide
variety of FUV spectral energy distributions for normal elliptical galaxies.
However, observations indicate that E galaxy FUV spectra are remarkably
uniform (Burstein et al.\ 1988\markcite{B88}; Brown, Ferguson, \& Davidsen
1995\markcite{BFD95}) and consistent with a highly bimodal HB distribution
(Ferguson 1994\markcite{F94}).  Thus, for most purposes we can model
the hot stellar population with only two evolutionary tracks, one for a
PAGB star and one for an EHB star.  We then adjust $M_{env}$ for the 
EHB star and the fraction of the population in each type of star to achieve
the best fit.  In \S 4.4 we discuss the effect of relaxing this
assumption of extreme bimodality.
 
\subsection{Abundance Anomalies in Evolved Stars}

It is not a straightforward exercise to determine the 
abundances in the stellar populations producing the UV
light from elliptical galaxies.  This is because 
the UV absorption line strengths in the spectra of evolved stars do not provide
a direct measure of their inherent abundances.
Numerous observations of HB stars and their 
progeny in our own Galaxy show that these stars exhibit abundance anomalies
(see Saffer \& Liebert 1995\markcite{SL95} and references therein). 
Helium is usually underabundant in sdB stars while overabundant in sdO
stars.  Carbon and silicon are usually underabundant in both
sdB and sdO stars, while the nitrogen abundance usually appears normal in 
both classes (Saffer \& Liebert 1995\markcite{SL95};
Lamontagne, Wesemael, \& Fontaine 1987\markcite{LWF87};
Michaud, Vauclair, \& Vauclair 1983\markcite{MVV83}).  
The FUV spectra of evolved stars observed with HUT show absorption lines that
are markedly different from solar abundances or even scaled solar abundances
(Brown, Ferguson, \& Davidsen 1996\markcite{BFD96}).

These abundance anomalies can be partly explained by 
models of diffusion processes in evolved stars, which demonstrate
that abundances can be greatly enhanced or diminished through chemical 
separation in the stellar envelope and atmosphere. 
Individual elements
can either levitate or sink in the outer layers of a star, depending
upon which force has a stronger effect on each element: gravity (g)
or radiative acceleration (g$_{\rm rad}$).  
In theory, the tendency is for heavy elements to sink in cooler stars, and for 
these elements to levitate in hotter stars.
If g$_{\rm rad}$ is larger than g,
an element will be radiatively supported and it accumulates
in the atmosphere, increasing
line saturation, which in turn will decrease g$_{\rm rad}$,
until an equilibrium can be reached between g and g$_{\rm rad}$
(Bergeron et al.\ 1988\markcite{BWMF88}).  If g$_{\rm rad}$ is less than g, the
element will sink into the star.  However, the resulting abundance decrease
may or may not lead to an increase in the local value of g$_{\rm rad}$, 
depending upon whether or not the lines were originally 
saturated or unsaturated.  Thus,
an equilibrium between g and g$_{\rm rad}$ may or may not
be reached (Bergeron et al.\ 1988\markcite{BWMF88}).
Michaud et al.\ (1983\markcite{MVV83}) demonstrated
that if the abundances of C, N, O, Ca, and Fe are all assumed to be
originally at 1/200 solar abundance, one can achieve abundance enhancements
of three orders of magnitude through diffusion on the timescale of 10$^8$ yr.
Unfortunately, diffusion theory has not reached the level of accuracy
needed to predict reliably the effects observed in evolved stars.
For example, Bergeron et al.\ (1988\markcite{BWMF88}) calculated
three models with effective temperatures of 20,000~K, 35,000~K, and 50,000~K,
and with respective gravities of log~g~=~5.0, 5.5, and 6.0, all
originally at solar abundance.   
They found that while the nitrogen abundance was
at approximately solar abundance for all models, agreeing with 
observations, the behavior of carbon and silicon did not agree with
observed trends.  
In the models, carbon was underabundant and increasing with
T$_{\rm eff}$, yet observations show that the carbon abundance seems to
decrease with increasing T$_{\rm eff}$ (Bergeron et al.\ 1988\markcite{BWMF88}
and references therein).
Silicon was predicted to be between 0.1 and 1 of the solar
value, but silicon deficiencies by factors of 10$^4$--10$^5$ are observed
for these stars.  Such discrepancies might point to other processes at
work, such as a weak stellar wind (Bergeron et al.\ 1988\markcite{BWMF88}).
The bottom line is that these diffusion processes greatly complicate
an absorption line analysis of an evolved population, and the strengths
of the absorption lines are unlikely to reflect the inherent abundance
of the population.  Since the stronger metallic lines may well be saturated in 
stellar envelopes with very high abundances, metals in these envelopes
may not be radiatively supported, allowing them to sink into the star.
At the other extreme, very low abundances can be enhanced by orders of
magnitude.  Thus, diffusion processes tend to enhance low intrinsic
abundances and diminish high intrinsic abundances.

In the elliptical galaxies,
line analysis is further complicated by the fact that 
the spectra represent composite systems
with components that cover a range in temperature and gravity, which
in turn affect line strengths just as abundance does.  Furthermore, in the
FUV, the density of lines is so large that only the very strongest lines
will be clearly distinct in these composite spectra.

With all of these complications in mind, an
absorption line analysis alone will not answer the
metallicity debate; we must look to 
the shape of the UV continuum as well.  The shape of the continuum
reflects the composite nature
of these galaxies which no single temperature stellar model may duplicate
(Brown et al.\ 1995\markcite{BFD95}), and thus reflects
the evolutionary paths of the individual stars in the composite model.
In this study, we use
the stellar evolutionary models of DRO93\markcite{DRO93},
Bressan et al.\ (1993\markcite{BFBC93}), and 
Fagotto et al.\ (1994a\markcite{FBBC94a}, 1994b\markcite{FBBC94b},
1994c\markcite{FBBC94a}) to construct synthetic spectra for
model stellar populations.
The individual synthetic spectra which comprise the composite models
are from the grids of Brown et al.\ (1996\markcite{BFD96})
and Kurucz (1992\markcite{K92}), for 
T$_{\rm eff} \geq$~10,000~K and T$_{\rm eff} <$~10,000~K respectively.  
The composite models 
which best fit the elliptical galaxy spectra draw the bulk of their flux from
stars in the range 15,000~K~$\leq$~T$_{\rm eff}$~$\leq$~30,000~K.

\section{OBSERVATIONS}

In March of 1995, the Hopkins Ultraviolet Telescope (HUT) was flown
aboard the Space Shuttle \it Endeavour \rm during the Astro-2 mission.
It collected FUV spectra of six elliptical and S0 galaxies;
these observations are summarized in Table 1.  Together
with the spectra of two galaxies from Astro-1,
these data represent the only FUV spectra
of early-type galaxies with wavelength coverage down to the Lyman limit
at 912~\AA, and thus include the ``turnover'' in the spectra that
is needed to fully characterize the spectral energy distribution.
As determined by 
Burstein et al.\ (1988\markcite{B88}) with IUE, the ($m_{1550}-V$) color
varies over a range 2.04--3.86~mag for these
galaxies.  All of them suffer from a minimal amount of reddening
($0.00\leq E(B-V)\leq 0.035$~mag).

HUT is uniquely suited for FUV spectroscopy of
nearby galaxies.
The telescope uses a 90~cm diameter
$f/2$ primary mirror and a prime focus, near-normal-incidence,
Rowland-circle grating spectrograph to obtain spectrophotometry
from 820 to 1840~\AA\ in first
order with a resolution of 2--4~\AA.  The detector consists of a
microchannel plate coupled to a phosphor intensifier that is read
by a Reticon into 2048 pixels.  
All six galaxies were observed during orbital 
night through a $10 \times 56 \arcsec$ slit.
A complete description of the
instrument and its calibration is given in Davidsen et al.\
(1992\markcite{DLD92}).   Modifications and the resulting performance
and calibration for the Astro-2 mission are described in
Kruk et al.\ (1995\markcite{K95}).  The final Astro-1 calibration of HUT
is presented in Kruk et al.\ (1996\markcite{K96}).
The fast focal ratio, large apertures (modified somewhat for Astro-2),
and wavelength coverage down to the Lyman limit make HUT an 
ideal instrument for
observing extended objects and for determining the effective
temperatures of hot UV sources.  
For Astro-2, an 
increase in the effective area by a factor of 2.3, combined with an
extended duration flight, allowed HUT to collect 
much more data with higher signal-to-noise (S/N) than on Astro-1. 

Fig.~2 shows the HUT slit superimposed on two different images
of the elliptical galaxy M~60 (NGC~4649) and the nearby spiral galaxy 
NGC~4647.   The top panel is an optical image 
($\sim$5850--6820~\AA; Minkowski \& Abell 1963\markcite{MA63})
from the Digitized Sky Survey (DSS), and the bottom panel is an FUV 
image (1344--1698~\AA) from the 
Ultraviolet Imaging Telescope (courtesy of UIT; see also Ohl \& O'Connell 
1997\markcite{OO97}).  
The figure demonstrates several
important points.
Of the six Astro-2 galaxies considered in this paper, M~60 is the only galaxy
that has a possible close companion.  However,
the pairing is likely due to a superposition on the sky:  although
the two galaxies have similar redshifts, the lack of evidence of
tidal interaction, either as tidal plumes or as a distortion of the spiral
structure in NGC~4647, suggests that the galaxies are at different distances
(Sandage \& Bedke 1994\markcite{SB94}).
The bottom panel shows that the FUV light is very concentrated in M~60,
and that a large percentage of this light falls within the HUT slit.
The FUV image also shows that although the FUV light from the spiral galaxy
is more extended than that in the elliptical galaxy, 
our HUT spectrum is clearly not contaminated
by the FUV flux from NGC~4647.

Our six spectra were corrected for airglow contamination from the 
lines Lyman~$\alpha$~$\lambda1216$, Lyman~$\beta$~$\lambda1026$, 
Lyman~$\gamma$~$\lambda973$, and
\ion{O}{1}~$\lambda1305$ by fitting airglow profiles
simultaneously with our model fitting.  The spectra were fluxed and
corrected for interstellar 
extinction in our Galaxy using the values of $E(B-V)$ from Table 1, 
$R_V$~=~3.05, and the extinction parameterization of 
Cardelli, Clayton, \& Mathis (1989\markcite{CCM89}). 
For the stellar population model-fitting described in this paper, 
the HUT data have been
binned such that each wavelength point represents 5 HUT bins or 2.5~\AA.
Only those portions of the observation with stable pointing and 
the galaxy well centered in the slit were used.
Fig.~3 shows the HUT spectra for each of the six Astro-2 galaxies, along with
models that will be described below.  The NGC~1399 Astro-1 data has been 
shown in several previous papers (e.g. Ferguson et al.\ 1991\markcite{F91}).

\section{EVOLUTIONARY MODELS}

In order to model the FUV luminosity from the various classes of HB
stars and their progeny, we construct synthetic composite spectra for 
evolved stellar populations by integrating 
individual synthetic stellar spectra over the stellar
evolutionary tracks of DRO93\markcite{DRO93}, 
Bressan et al.\ (1993\markcite{BFBC93}), 
Fagotto et al.\ (1994a\markcite{FBBC94a}, 1994b\markcite{FBBC94b},
1994c\markcite{FBBC94a}), and Vassiliadis \& Wood (1994\markcite{VW94}).
The synthetic stellar spectra come from the grid of Brown et al.\ 
(1996\markcite{BFD96}) for effective temperature T$_{\rm eff}\geq10,000$~K,
and from the grid of Kurucz (1992\markcite{K92}) for T$_{\rm eff}<10,000$~K
(although these stars below 10,000~K do not significantly contribute 
to the integrated FUV luminosity).
At each point along an evolutionary track, we find
the synthetic stellar spectrum closest in T$_{\rm eff}$ and 
surface gravity (g).  To correct for the small mismatch in T$_{\rm eff}$ and g,
we scale the bolometric luminosity from the
synthetic spectrum so that it matches that in the track step.  We then 
weight the contribution from that point by the time spent at that
location in the HR diagram.  Our composite spectrum for a given track
thus corresponds to a population that enters that track at the rate
of one star per year.  The normalization of the model and the
distance to the galaxy then determine the actual rate of stars entering
and leaving the evolutionary tracks, or the stellar evolutionary flux (SEF).
Examples of these integrated composite
spectra are shown in Fig.~1, for stars evolving along PAGB, PEAGB, and AGBM
evolutionary paths.  

The diffusion processes in the outer layers of these evolved stars create
inconsistencies between the abundances that determine
the evolution of the stars and the abundances in the stellar atmospheres
of these stars (see \S 1.2).
Because of these inconsistencies, we consider synthetic spectra with three
different atmospheric metallicities ($Z = Z_{\sun}$, 
$Z = 0.1$~$Z_{\sun}$, and $Z = 0.01$~$Z_{\sun}$)
for all of the different evolutionary paths, regardless of the inherent
abundances driving that evolutionary path.  
We define $Z_{atm}$ as the abundance of
heavy elements driving the \it line strengths \rm in our composite 
synthetic spectra, and we define $Z_{evol}$ as the abundance of
heavy elements driving the \it evolution \rm of stars along the 
evolutionary path in a given composite synthetic spectrum.

The limitations of our modeling scheme are the following: the small
mismatch in temperature and surface gravity between the individual
synthetic stellar spectra and the evolutionary track points; the
assumption of local thermodynamic equilibrium in the synthetic
spectra with T$_{\rm eff}$~$<$~45,000~K; pure H \& He atmospheres for 
T$_{\rm eff}\geq$~20,000~K (with metals incorporated only in computing
the emitted spectrum, not the atmospheric structures); 
the lack of continuum opacity for 
elements heavier than helium for spectra with 
T$_{\rm eff}\geq$~20,000~K; and the
omission of mass loss during the HB and AGB phases in the 
DRO93\markcite{DRO93} evolutionary tracks, as well
as the usual
uncertainties and simplifying assumptions that go into stellar
evolution calculations.  Errors arising from the tracks are discussed
in DRO93\markcite{DRO93} and in \S 4.3.  
These limitations may affect the quantitative results, but
are less likely to affect the qualitative conclusions.

\section{MODEL FITTING}

As discussed in the previous section, the normalization of a model
fit to the HUT data gives the stellar evolutionary flux (SEF) for stars 
evolving along the evolutionary
path for that model (in the HUT slit).  We also have a
theoretical value of the SEF$_{\rm tot}$~= 
the rate of stars evolving along \it all \rm evolutionary paths in the galaxy.
For old stellar populations (age~$\geq 10^{10}$~yr), 
the SEF$_{\rm tot}$ per unit luminosity is approximately 
2.2$\times$10$^{-11}$ stars
L$_{\sun}^{-1}$ yr$^{-1}$ (Greggio \& Renzini 1990\markcite{GR90}).
There is a dependence of this quantity on composition, particularly
on the He abundance, since helium-rich stars at a given mass evolve
more quickly.  Our models assume that in a galaxy, the dominant population
that controls the effective value of the SEF$_{\rm tot}$ has
$Y$ close enough to $Y_{\sun}$ 
such that the solar value of the SEF$_{\rm tot}$
is appropriate. 
To determine the bolometric luminosity for each of the Astro-2 galaxies, 
we calculate the $V$-magnitudes 
through the HUT slit from archival HST WFPC images, correcting the 
0.06~mag color offset between the F555W filter and the Johnson $V$ band and 
using a bolometric 
correction (BC) of --1.32.  This BC is consistent with a stellar
population of age 12~Gyr and [Fe/H]~=~0.25 (Worthey 1992\markcite{W92}).
For a given evolutionary model fit to the HUT data, normalizing our
measured SEF by the bolometric luminosity gives a distance-independent 
measurement of the fraction of stars evolving along that evolutionary path,
since we know the SEF$_{\rm tot}$.
 
For each of the six Astro-2 galaxies, we perform two different classes of
fitting: EHB stars alone, or EHB and PAGB stars together.  
In all of our fitting,
we perform $\chi^2$ minimization with the
IRAF routine SPECFIT (Kriss 1994\markcite{K94}).
For every model fit to the data, we assume that neutral hydrogen from the
interstellar medium (ISM) absorbs at a column depth equivalent to that
along a line of sight in our Galaxy, as listed in Table~1.

\subsection{EHB tracks}

We initially fit EHB composite models based
on evolutionary tracks from a variety of groups
(DRO93\markcite{DRO93};
Bressan et al.\ 1993\markcite{BFBC93};  
Fagotto et al.\ 1994a\markcite{FBBC94a}, 1994b\markcite{FBBC94b},
1994c\markcite{FBBC94a}).  
While EHB/AGBM sequences from the latter two groups were found to 
produce reasonable fits to the HUT data, the DRO93\markcite{DRO93} tracks were 
found to have somewhat superior fits. 
In addition, they provide a finer mass spacing and the ability to
check the effects of $Y_{ZAMS}$ and $Z_{evol}$ separately (as well
as the fact that we can 
compute additional physically consistent sequences where
necessary).  At fixed chemical composition and envelope mass,
spectra constructed with the Fagotto et al.\ (1994a, 1994b, 1994c)
and Bressan et al.\ (1993\markcite{BFBC93}) tracks agree with those
constructed from the DRO93\markcite{DRO93} tracks.
Thus, to simplify
the discussion, we limit our 
analysis to the DRO93\markcite{DRO93} tracks.

The DRO93\markcite{DRO93} tracks are broken into subsets
according to the abundances of metals and helium (see their Table 1 and 
our Table 2); there are eight different abundance combinations.
The tracks detail the evolution of HB stars from the zero-age horizontal
branch through core He exhaustion, and continue either to
the thermally pulsing stage (for AGB evolution) or to the white dwarf
cooling track (for AGBM and PEAGB evolution).  Each track is
parameterized by abundance, ZAHB core mass, and ZAHB envelope mass.  
We took all 136 tracks in their paper and constructed a corresponding 
composite spectrum for each track, which was then fit to the HUT data.
These spectra were computed with metallicities of 
$Z_{atm}$~=~$Z_{\sun}$, 
0.1~$Z_{\sun}$, 0.01~$Z_{\sun}$.  We found that those tracks with
$Z_{atm}$~=~0.1~$Z_{\sun}$ best matched the overall absorption features in
the HUT data (see \S 6).  The best fit evolutionary model
(with $Z_{atm}$~=~0.1~$Z_{\sun}$)
for each of the Astro-2 galaxies is listed in Table~3 and plotted in Fig.~3.
All of the best fits come from two track sets 
with high $Z_{evol}$ and $Y_{ZAMS}$ (sets G \& H in Table 2). 
We also include results of the same fitting applied to the spectrum of 
NGC~1399, which was observed through a $9.4 \times 116 \arcsec$ slit
on Astro-1 (we omit the M~31 Astro-1 data because of the large and uncertain
extinction toward that galaxy).  The results in Table 3 show that the
SEF for EHB stars (SEF$_{\rm EHB}$) 
is a small fraction ($\lesssim 10$~\%) of the 
SEF$_{\rm tot}$.  In a model comprised solely of EHB stars, one
can reproduce the amount of FUV light from these galaxies without
approaching the limits imposed by the fuel consumption theorem, since
EHB stars and their progeny are such efficient FUV emitters.
Using the $\chi^2$ statistics to judge the quality of fit, we can compare
the quality of the best-fitting EHB model from each abundance
group (cf.\ Table 2) to that of the best fit of any group.  
We make this comparison in Table 3, which shows a strong preference
for groups G and H in the fitting.

We shall use the FUV spectrum of M~60 to demonstrate how models with higher
metallicity and helium abundance produce a closer match to the HUT data,
since our M~60 spectrum has the best S/N ratio.
Fig.~4 shows the best fit $Z_{atm}$~=~0.1~$Z_{\sun}$ model 
for each of the 8 different abundance 
($Z_{evol}$) sets.
In the left column are the evolutionary models for stars with abundances
less than or equal to the solar abundance (sets A--D in Table 2).  These plots
show a significant deficit in the flux at the shortest wavelengths in the 
model, as compared to the HUT data for M~60.  
On the other hand, the evolutionary
models in the right column, for stars evolving at higher metal and helium
abundance (sets E--H in Table 2), 
show a somewhat smaller flux deficit at the shortest wavelengths.
The evolutionary model that fits best is that with $Z_{evol}$~=~0.04 and 
$Y_{ZAMS}$~=~0.34 (set G).

If we define the flux deficit as

\[\Delta F = \frac{\displaystyle \sum_{\lambda=912 \AA}^{970 \AA} (F_{data} - F_{model})}{\displaystyle \sum_{\lambda=912 \AA}^{970 \AA}F_{data}}\]

\noindent
we can characterize the deficit numerically.  These deficits (listed in 
Fig.~4) clearly show that the low $Y_{ZAMS}$ and $Z_{evol}$ models in the 
left column are more deficient in this wavelength range,
as compared to the high $Y_{ZAMS}$ and $Z_{evol}$ models in the right 
column.  However, even the models in the right column show a significant
flux deficit.  This deficit could be due to the absence of flux from PAGB 
stars in these models.
Because of the high integrated luminosity for stars evolving along 
EHB tracks (see Fig.~1), 
the SEF$_{\rm EHB}$ is a small fraction of the
rate of all stars evolving along any evolutionary path in these galaxies
(the SEF$_{\rm tot}$; cf. Table 3).  
Most of the stars must be evolving along the classical
PAGB tracks. 

The low-metallicity models appear to have a stronger flux
deficit than the high-metallicity models; the reason may be related to the 
ratio of flux emitted in the HB and post-HB phases. We note that
the mean temperature of the integrated spectrum for a given evolutionary
sequence depends on the ratio of the radiation emitted in the post-HB
stages to that emitted before core He exhaustion, simply because the
mean temperature of AGBM stars is higher than the earlier stage. 
If one takes the integrated flux below 1200~\AA, and computes the 
ratio of this flux in the post-HB phase to that in the HB phase, one
finds that this ratio increases as $Y_{ZAMS}$ and $Z_{evol}$ increases.
This is 
at least a partial explanation why our models tend to favor the higher
metallicities.  

The models that work best from any of the abundance ($Z_{evol}$ \& $Y_{ZAMS}$)
sets and for any $Z_{atm}$ 
are always those that integrate over AGBM evolutionary tracks 
(cf. Fig.~1).
Models that integrate over PEAGB tracks have spectra that are too cool to
agree with the HUT data.  We find that a narrow range in envelope mass on
the EHB produces spectra that agree reasonably well with the HUT data, 
consistent with the findings in our previous analysis of these galaxies
(Brown et al.\ 1995\markcite{BFD95}).  This narrow distribution is on the
blue end of the HB with low (but not the lowest) envelope masses on the HB.
We discuss the distribution below in the context of the more realistic
two-component EHB + PAGB models.

\subsection{EHB and PAGB tracks}

Because a very small fraction 
of the stars are evolving along EHB tracks in these
galaxies, a more appropriate model to fit to the data is a two-component 
model, with both EHB stars and PAGB stars.  Our PAGB models are constructed
by integrating over the 
six solar-metallicity H-burning tracks of Vassiliadis \& Wood 
(1994\markcite{VW94}) with core masses $M_{core}^{PAGB}$=0.569, 0.597, 0.633,
0.677, 0.754, and 0.900 $M_{\sun}$.  Vassiliadis \& Wood 
(1994\markcite{VW94}) also calculate H-burning tracks at lower metallicities
(six at $Z$~=~0.008, four at $Z$~=~0.004, and two at $Z$~=~0.001) and a 
handful of He-burning tracks.  Since all of the H-burning tracks appear
qualitatively similar, and since He-burning PAGB evolution is thought to
be in the minority ($\lesssim$ 25\%; Vassiliadis \& Wood 
1994\markcite{VW94} and references therein), 
we take the six H-burning solar-metallicity tracks as representative 
of the evolution for classical PAGB stars.
We note that all of the Vassiliadis
\& Wood (1994\markcite{VW94}) tracks start at T$_{\rm eff}$~=~10,000~K in the 
HR diagram, after the thermally pulsing (TP) stage (see Fig.~1).  Our
PAGB model spectra thus omit the flux that is generated as the star evolves
along the AGB from the HB to the TP stage, since
in the HUT wavelength range the contribution from the cool AGB is negligible.

We used two methods to fit two-component (EHB \& PAGB) models to the HUT data.
The first method allowed the contributions of EHB stars and PAGB stars to
float both freely in the fit.  The second method constrained the contribution
of the PAGB component so that it would not exceed the SEF$_{\rm tot}$.
Assuming that the SEF$_{\rm tot} \approx 2.2 \times 10^{-11}$ stars
$L_{\sun}^{-1}$ yr$^{-1}$ $\pm$10\% (Greggio \& Renzini 1990\markcite{GR90}),
and that EHB stars by themselves can
at most contribute a small fraction ($\lesssim$10\%) of the SEF$_{\rm tot}$
(from our own fitting in the previous section, and from Brown et al.\
1995\markcite{BFD95}), we restricted 
only the PAGB component of the fit, 
and not the sum of the EHB and PAGB components.
Within the uncertainty of the SEF$_{\rm tot}$, the EHB contribution
to the SEF$_{\rm tot}$ is small.
This assumption simplifies the $\chi^2$ minimization.

When we did not restrict the PAGB contribution to the model fit, we found that 
EHB stars from all eight of the DRO93\markcite{DRO93}
metallicity groups could reasonably fit the HUT data;
the fitting only slightly favored the high $Z_{evol}$ and high $Y_{ZAMS}$
models (sets E--H in Table 2).  This was because the 
short-wavelength deficiencies of the models with $Z_{evol}$~$<$~$Z_{\sun}$
were compensated by a large
PAGB contribution ($\approx$ half of the FUV flux).  However,
these solutions were not astrophysically consistent,
since the PAGB contribution allowed in the fits was much
larger than that allowed by the theoretical SEF$_{\rm tot}$ (unless our
models severely underestimate the potential UV flux from PAGB stars).

When we restricted the PAGB contribution so that it could not exceed
the theoretical SEF$_{\rm tot}$, we found that the EHB evolutionary
models with high $Z_{evol}$ and high $Y_{ZAMS}$ 
were favored in the fits to the
HUT data.  The strength of this result is a function of the 
assumed $M_{core}^{PAGB}$
in the galaxies, in the sense that the high $Y_{ZAMS}$ and high
$Z_{evol}$ models are more strongly favored as the 
assumed $M_{core}^{PAGB}$ increases.  
Our results depend upon $M_{core}^{PAGB}$ because
as the mass increases, the
PAGB stars evolve more rapidly and produce 
less FUV flux over their lifetimes.
Thus, if the stellar evolutionary flux
of PAGB stars (SEF$_{\rm PAGB}$) for each $M_{core}^{PAGB}$
is limited by the same SEF$_{\rm tot}$,
the more massive PAGB stars will contribute less flux to the EHB + PAGB
models than less massive PAGB stars.  

Since the strength of our results depends upon the $M_{core}^{PAGB}$
assumed, we show these results for the two lowest masses in the
set of PAGB models: those with
$M_{core}^{PAGB}$~=~0.569~$M_{\sun}$ (Table 4 and Fig.~5) and 
0.597 $M_{\sun}$ (Table 5).  
Our results are stronger as $M_{core}^{PAGB}$ increases, and thus
the results in Table 4 represent the conservative results, while
those in Table 5 use a $M_{core}^{PAGB}$ that is closer to that
modelled in the planetary nebula luminosity functions of 
ellipticals (0.61~$M_{\sun}$; see below).  
In these EHB + PAGB models,
the SEF$_{\rm EHB}$ is a small fraction of the SEF$_{\rm tot}$
predicted by the fuel consumption theorem, while the SEF$_{\rm PAGB}$ is
much larger.  Note that consistent models only require that 
$\rm SEF_{PAGB} \leq SEF_{tot}$,
and not $\rm SEF_{EHB} + SEF_{PAGB} = SEF_{tot}$.  The value of
SEF$_{\rm PAGB}$ that we derive depends on the integrated flux of a particular
assumed PAGB track.  In reality the PAGB population will have a mass
spectrum whose mean mass can be higher and it integrated flux therefore
lower, in which case our low-mass PAGB
models will underestimate the number of PAGB
stars present while overestimating the mean UV contribution from
individual stars.
The individual PAGB tracks have very
similar integrated spectra, and so we cannot constrain the PAGB mass
with our fitting.  

Using the $\chi^2$ statistics to judge the quality of fit, we compare
in Tables 4 and 5
the quality of the best-fitting EHB + PAGB models from each abundance
group (cf.\ Table 2) to that of the best fit.  
In Table 4, evolutionary abundances
of at least solar are favored in 5 of the 7 galaxies.  The exceptions
are NGC~3115 and NGC~3379, where there are no preferences for either high
or low evolutionary
abundances in the fitting.  Alternatively, in Table 5,
evolutionary abundances greater than the
solar value are favored in all 7 of the galaxies, with most fits favoring
high $Z_{evol}$ and $Y_{evol}$.  When $M_{core}^{PAGB}$ was increased
to even higher values, the fits for all of the galaxies 
increasingly favored higher $Z_{evol}$
and $Y_{ZAMS}$.  The culmination of this trend was shown in the previous
section, where we fit models consisting solely of EHB stars.  Such
models are equivalent to the assumption that the PAGB stars are too 
massive to contribute significant flux to the FUV spectra.

What is the appropriate $M_{core}^{PAGB}$ to use in our fitting?
Studies of the planetary nebula luminosity function (PNLF) may indicate that
$M_{core}^{PAGB} \approx$~0.61~$M_{\sun}$.
The PNLF is well-studied in 
both spiral bulges and elliptical galaxies.  Ciardullo et al.\ 
(1989\markcite{CJFN89}) calibrated the PNLF in the bulge of M~31,
and found that it was in excellent agreement with a set of PAGB tracks
having a  mean $M_{core}^{PAGB}$ of 0.61~$M_{\sun}$.  Since then, the same PNLF
has demonstrated very good agreement with that in the bulge of the
Sb galaxy M~81 (Jacoby et al.\ 1989\markcite{JCFB89}), in E and S0 galaxies
of the Leo I Group (Ciardullo, Jacoby, \& Ford 
et al.\ 1989\markcite{CJF89}), and in E and S0 galaxies of the Virgo Cluster 
(Jacoby, Ciardullo, \& Ford 1990\markcite{JCF90}).  Four of the Astro-2
galaxies (M~49, M~60, M~87, and NGC~3379) were included in these PNLF
studies.   In the inner regions (isophotal radius $<$~$1\arcmin$) of
the Virgo and Leo I ellipticals, PN were lost in the bright galaxy background,
but no significant gradients in the 
luminosity-specific PN density were seen beyond these regions.  Thus,
we can not know for certain the characteristics of the PAGB population
within the cores of the ellipticals, but the evidence we have
points to $M_{core}^{PAGB} \approx$~0.61~$M_{\sun}$.  

In the literature there are other sets of PAGB tracks that extend the range of
core masses even lower than the 0.569~$M_{\sun}$ track of 
Vassiliadis \& Wood (1994\markcite{VW94}) and thus increase the potential
PAGB contribution to the total UV flux.  Although it is unlikely that
all of the PAGB stars in a galaxy would be evolving along such low-mass tracks
(see above), we investigated the effect this would have on our modeling.
For example,
the Sch$\ddot{\rm o}$nberner (1987\markcite{S87}) set has a track with a core
mass of 0.546~$M_{\sun}$.  Such a low-mass PAGB star 
evolves so slowly that it contributes
significant flux as the PAGB component of
an EHB + PAGB model, even when the Sch$\ddot{\rm o}$berner track is
restricted by the SEF$_{\rm tot}$ (although we stress
that the integrated flux from this 0.546~$M_{\sun}$ track does not fit
the HUT data by itself, without an EHB/AGBM component).
So, EHB + PAGB models that used a 0.546~$M_{\sun}$
PAGB track did not show a preference for either the high $Z_{evol}$
or low $Z_{evol}$ models; the results were similar to those 
of the unconstrained EHB + PAGB models above.
However, Sch$\ddot{\rm o}$nberner states
that this 0.546~$M_{\sun}$ model is below the threshold for the
occurrence of thermal pulses; the model is therefore actually a PEAGB
sequence.  Although pre-TP evolution 
contributes negligibly to the integrated FUV luminosity of stars
evolving along PAGB tracks, the same cannot be said of all PEAGB stars.  If
we include the luminosity from a 0.550~$M_{\sun}$ AGB track (DRO93) 
as a plausible example of the pre-TP evolution for
a star evolving along the 0.546~$M_{\sun}$ Sch$\ddot{\rm o}$nberner track,
we find that the composite spectrum appears similar to the PAGB spectrum
in Fig.~1, except that the spectrum rises considerably at wavelengths longer
than 1500~\AA, and thus does not agree with the 
Astro-2 data.  The result is that this ``PAGB'' contribution (which is really
PEAGB) in an EHB + PAGB fit to the data is minimized.  In summary, 
the 0.546~$M_{\sun}$ track is an acceptable PAGB component only 
if two conditions are met.  First, all of
the PAGB stars in the galaxies must evolve along extremely low-mass
tracks.  Second, the pre-TP evolution must be characterized by high mass stars
that are too cool to produce flux in the HUT range while on the AGB, 
yet lose enough mass on the AGB to 
subsequently become 0.546~$M_{\sun}$ PAGB stars.

Because the ``pseudo-continuum'' 
shape is affected by $Z_{atm}$, the best fitting model was usually, but
not always, that with $Z_{atm}$~=~0.1~$Z_{\sun}$.  However,
at any $Z_{atm}$, the fitting still favored
those evolutionary models with high $Z_{evol}$ and high $Y_{ZAMS}$.  Since
the individual
absorption features are better matched by $Z_{atm}$~=~0.1~$Z_{\sun}$
(see \S 6), we
chose to demonstrate the best fits to the HUT data by consistently
using models with this atmospheric abundance.

Although the models in Fig.~5 fit the Astro-2 data fairly well, there are
a few curious discrepancies.  The M~49 data shows a large
excess at 961~\AA, although the deviation from the model is not
very significant.  The formal probability of such a deviation
from the model flux is 0.16.  However, a more significant discrepancy appears
in the NGC~3379 data as an emission feature
near \ion{C}{4}$\lambda\lambda1548,1551$.  
The feature can be fit by a Gaussian with $\lambda$~=~1562.7~\AA, 
equivalent width =~10~\AA, and FWHM~=~3000~km~$s^{-1}$.  The statistical
significance of the feature is 0.0084 (better than a 2$\sigma$ detection).
But what could be causing this feature?  
We observed NGC~3379
twice, and the same feature shows up in both observations, so it is unlikely
that some ``glitch'' in the data is causing the feature.
The deredshifted wavelength of
the feature would be 1557.9~\AA, which is still 7~\AA\ longward of the
most likely emission at \ion{C}{4}.  If an extraneous
emission source fell within the HUT slit, such as a supernova remnant or
cooling flow, one would expect to find other emission lines 
(Ly$\alpha$, \ion{O}{6}, etc.) at the same redshift.  Given the lack
of other detectable emission features in the NGC~3379 spectrum, we tentatively
conclude the feature is a fluke; a 2$\sigma$ fluke is not unreasonable
given 6 data sets having 410 bins each.

The best-fitting two-component models incorporate EHB tracks with 
a narrow range in temperature and 
envelope mass on the EHB, but this range does not include the very 
lowest envelope masses, which extend down to 0.002 $M_{\sun}$.  
Since the $\chi^2$ contours are probably dominated
by the systematic uncertainties of the models,
we conservatively use
the $\chi^2$ statistics to determine the range of $M_{env}$ that
is within 4$\sigma$ of the best-fit model.  We find that in our
two-component EHB + PAGB models, a narrow
distribution of $M_{env}$ on the EHB fits the Astro-2 data 
within 4$\sigma$, as we found in our initial analysis of these data
using lower resolution (10~\AA) models 
(Brown et al.\ 1995\markcite{95}).  The range for each galaxy is listed in
Tables 4 and 5.

As in the previous section with the one-component (EHB) models, we
use the high S/N data for M~60 to demonstrate why models with high
$Z_{evol}$ and high $Y_{ZAMS}$ work better than models
with low $Z_{evol}$ and low $Y_{ZAMS}$.
Fig.~6 shows the best fit
EHB + PAGB model from Table 4, 
for each of the 8 different abundance 
($Z_{evol}$) sets.
In each panel, the PAGB component has $M_{core}$~=~0.569~$M_{\sun}$.
The flux deficiency ($\Delta$F) at the shortest wavelengths is more
significant in the low $Z_{evol}$ and low $Y_{ZAMS}$ models than in the
high $Z_{evol}$ and high $Y_{ZAMS}$ models.  However, the high
$Z_{evol}$ \& $Y_{ZAMS}$ models do still show some deficiency at the shortest
wavelengths, and we discuss this problem in the next section.

In our previous study of these galaxies
(Brown et al.\ 1995\markcite{BFD95}), we found that PAGB stars
alone do not fit the FUV spectral energy distribution of any of the 
galaxies.  This result also holds true in the current study.
The result that none of the galaxies can be fit by PAGB stars is
significant, because if one were trying to reproduce only the flux
at 1550~\AA, and not the entire FUV spectral energy distribution
observed with HUT,
galaxies with very red ($m_{1550} - V$) colors could be 
accounted for by PAGB stars alone.
Hence all of the galaxies in our study appear to require some
EHB fraction.

\subsection{Systematic Errors}

As we discussed above, HB models from DRO93\markcite{DRO93}
with supersolar $Y_{ZAMS}$ and $Z_{evol}$ 
fit the HUT data 
better than models with subsolar $Y_{ZAMS}$ and $Z_{evol}$.  This is true
for both EHB models and EHB + PAGB models. 
Although the supersolar $Y_{ZAMS}$ and $Z_{evol}$
models work better across the entire HUT range, the largest improvement is
in the region shortward of 970~\AA.  But how much can systematic errors
contribute to this result?  
The major sources of systematic error in our modeling are errors
in the individual stellar synthetic spectra, uncertainty in the 
extinction correction for the galaxies, and uncertainties in
the behavior of the input stellar evolution models.

As described in Brown et al.\ (1996\markcite{BFD96}), the Lyman series lines
are somewhat 
too strong in our grid of synthetic stellar spectra.  This produces a
deficiency in the synthetic spectra as one moves to shorter wavelengths
in the HUT range.  In Brown et al.\ (1996\markcite{BFD96}), 
there are six plotted synthetic stellar spectra that have flux down to the
Lyman limit, at effective temperatures of 17000~K, 24000~K, 29900~K, 36100~K,
40000~K, and 55000~K, and each has HUT data or a Kurucz LTE model for 
comparison.  If we measure the flux deficiency $\Delta$F for these 
6 models, we find that the average deficiency is 0.12$\pm$0.08.
Therefore, 
the supersolar $Y_{ZAMS}$ \& $Z_{evol}$ models in the right column of
Fig.~6 have deficiencies that fall within the range expected from 
systematic errors in our individual synthetic spectra, whereas the 
subsolar $Y_{ZAMS}$ \& $Z_{evol}$ models in the left column have 
deficiencies that fall outside of that range.

The extinction parameterization of Cardelli et al.\ (1989\markcite{CCM89})
does not extend below 1000~\AA, but we have extrapolated it to the Lyman
limit at 912~\AA.  Although this results in some uncertainty in the
extinction correction, we note that our extinctions are quite low (see
Table 1), and thus any errors in the shape of the extinction curve will
in turn be quite small.  For example, our use of $R_V$~=~3.05, characteristic
of the very diffuse ISM, produces an extinction correction that deviates
from the galactic mean curve ($R_V$~=~3.1) by less than 1\% when 
$E(B-V)$~=~0.035, which is the highest extinction in Table 1.  The significant
uncertainty in the extinction is the possibility of extinction intrinsic
to the galaxies.  However, we note that if the Astro-2 galaxies have
internal extinction, our correction for this extinction would increase
the flux deficits for all of the models, and thus our supersolar $Y_{ZAMS}$ \&
$Z_{evol}$ models would still match the Astro-2 data better than those models
with subsolar $Y_{ZAMS}$ \& $Z_{evol}$.

The DRO90\markcite{DRO93} tracks neglect the effects of mass loss.
If mass loss takes place by winds during the EHB evolution, it
is likely to affect the brighter, low-gravity AGBM stage.
It will drive the evolution to higher temperatures
faster, but since it reduces the available hydrogen fuel
it is also likely to reduce the total UV output in these
stages.   The neglect of mass loss in the tracks should not affect 
our conclusions regarding metallicity.

\subsection{Mass Distribution on the HB}

The composite spectrum for a given evolutionary track is the
flux one would observe from a population of stars evolving from a single
mass on the HB.  In reality, the FUV spectra in elliptical galaxies 
probably arise from populations on the HB (and their progeny) with some
distribution of mass.  Unfortunately, the true mass distribution on the HB and
the mechanisms that govern it remain highly uncertain.  These uncertainties
are compounded by the
uncertainties in the age and composition of these stellar populations.  Thus,
there is scant theoretical justification for any given mass distribution.

As an initial attempt to constrain this problem,
we tried to fit Gaussian distributions of 
envelope mass (using a single core mass) to construct our populations,
but the best fit always occurred for a $\delta$-function in 
envelope mass, with the quality of the fits rapidly declining as the
width of the Gaussian increased.  We believed that a grid of evolutionary
tracks with a finer distribution in envelope mass would allow a Gaussian
distribution to fit, so we reran the evolutionary code of 
DRO93\markcite{DRO93} to create a new subset of tracks.
These 11 AGBM tracks were evenly spaced 
by $\Delta M_{env}$~=~0.002~$M_{\sun}$ 
around the best fit EHB track for M~60 in Table 3
($M_{env}$~=~0.036~$M_{\sun}$).

We then performed another two-component EHB + PAGB fit to the M~60 HUT data,
using the same $Z_{atm}$ and $Z_{evol}$ as the best EHB + PAGB fit in 
Table~5, but now the EHB contribution came from a Gaussian distribution
of envelope mass on the HB, such that the number of stars entering bins of
unit envelope mass were constrained by a symmetric Gaussian.  We found that
a narrow distribution of $M_{env}$,
centered on $M_{env}$~=~0.042~$M_{\sun}$ with FWHM~=~0.003 $M_{\sun}$,
fit the M~60 data slightly better 
than the original $\delta$-function at
$M_{env}$~=~0.046~$M_{\sun}$.
While this result suggests that the HB distribution is extremely bimodal,
the systematic uncertainties in the model atmospheres and the evolutionary
tracks combine to make the constraints weaker than they appear from formal
$\chi^2$ fitting.  Model spectra constrained with a fairly flat HB distribution
are not very different from those with a pure AGBM component, because
the long lifetimes make the AGBM stars dominant.  
For example, increasing the envelope mass FWHM from 0.003~$M_{\sun}$ to
0.1~$M_{\sun}$ increases $\chi^2$ by 20, and increasing the FWHM to
0.5~$M_{\sun}$ increases $\chi^2$ by 32, but the SED in these models
still provides a reasonable match to the continuum shape of the HUT data.  
Thus a ``flatter'' distribution in {\it mass} can produce a reasonable 
SED because
the distribution in {\it flux contribution} is still highly peaked around the
AGBM stars.  The 4$\sigma$ limits on the $M_{env}$ distribution given
in Tables 5 and 6 denote the limits on which stars can {\it dominate} the SED.

\section{COLOR-COLOR ANALYSIS}

The ($m_{1550}-V$) color of elliptical galaxies is the standard measure
of the strength of the UV upturn.
If galaxies with stronger
UV upturns
have a larger fraction of their stellar populations evolving
along EHB evolutionary paths than galaxies with weaker UV upturns, 
it then 
follows that galaxies with stronger UV upturns should also have a smaller
fraction of stars evolving along the classical PAGB tracks
(see Brown et al.\ 1995\markcite{BFD95} and 
Ferguson \& Davidsen 1993\markcite{FD93}).  
While the ($m_{1550}-V$) color tracks the strength of the FUV flux relative
to the optical flux, the ($m_{<1000}-m_{1550}$) color
tracks the ``characteristic''
temperature of the FUV population.
Using the HUT data and archival HST images,
we define these colors as

\[(m_{<1000}-m_{1550}) = 2.5~log_{10}~ \frac{<f_{\lambda=1450-1650 \AA}>}{<f_{\lambda=912-1000 \AA}>}\] and

\[(m_{1550}-V) = 2.5~log_{10}~ \frac{f_{V_o}}{<f_{\lambda=1450-1650 \AA}>}\]

\noindent
where log $f_{V_o=10}$ = --12.41 ergs cm$^{-2}$ s$^{-1}$ \AA$^{-1}$
(Johnson 1966\markcite{J66}), and
$V_o$ is derived from the $V$ and $E(B-V)$ values in Table 1. 

The left-hand panel of Fig.~7 shows the color-color diagram for the six
Astro-2 galaxies and for NGC~1399 (we again omit the Astro-1 galaxy
M~31 because of the large and uncertain extinction toward that galaxy).
The dashed line shows the shift in 
$(m_{1550}-V)$ that would occur if we used the measurements of Burstein
al.\ (1988\markcite{B88}) through the smaller IUE apertures.  
The large extent
and variation in the dashed lines is a clear indication of the 
strong variation in the 
FUV/optical flux ratio as a function of radius in these galaxies.
These differences in color gradients are apparent in 
the simultaneous UIT observations of these
galaxies (O'Connell et al.\ 1992\markcite{O92};
Ohl \& O'Connell 1997\markcite{OO97}).

In the right-hand panel, we show the variation in ($m_{<1000}-m_{1550}$)
for the integrated spectra of
models in abundance group G (Table 2).  The four reddest 
of these models (DRO93\markcite{DRO93}) stop at the TP stage (and
produce negligible FUV flux), so we
have tacked on plausible PAGB evolution from Vassiliadis \& Wood 
(1994\markcite{VW94}).
The ($m_{<1000}-m_{1550}$)
color appears ``hot'' for the high and low extremes of HB mass
(respectively HB/PAGB and EHB/AGBM). 
As one moves
from the blue end of the HB to larger envelope masses, the flux
from the EHB stars and their progeny becomes progressively
``cooler'' in the FUV.  This trend may explain the positions of the HUT data
points in the left-hand panel.  The hot NGC~1399 spectrum may be dominated
by stars evolving from the blue extreme of the HB, while the hot
FUV spectra of M~49, NGC~3115, and NGC~3379 may have significant contribution
from PAGB stars.  The ``cooler'' FUV spectra
of M~60 and M~89 may be dominated by AGBM stars with higher envelope masses
than those in NGC~1399.  Of course, with only a handful of galaxies,
it is premature to deduce a correlation between the behavior seen in the
data and that shown in the right-hand panel of Fig.~7.  But this type
of color-color diagram might be a useful tool in future studies.

M~87, which has an active nucleus, might have other reasons for its position
in the color-color diagram.  The active galaxies in the Burstein 
et al.\ (1988) study did not follow the UV-Mg$_2$ relation of the
quiescent galaxies.
However, recent HST observations
have shown that the FUV flux from the point source and jet in the M~87 nucleus
cannot be contributing significantly to the HUT spectrum (Tsvetanov
et al.\ 1996\markcite{TF96}).   
Also, subtracting the nuclear flux from the V measurement 
through the HUT slit would only diminish V and make the ($m_{1550}-V$)
color even bluer.  So why is M~87 so blue in both of the colors in Fig.~7? 
We believe that the answer may be related to the position of M~87 in the 
Virgo cluster.  Because M~87 lies at the center of this cluster, it has 
been accreting gas and metals from other galaxies.  Perhaps this
accretion has altered the stellar evolution in M~87 so that it is markedly
different from that in the other Astro-2 galaxies.  For example, 
a recent episode of
star formation could dramatically decrease ($m_{<1000}-m_{1550}$) but
not affect ($m_{1550}-V$) significantly.  If we start with
the best fit EHB + PAGB model for M~60, and add a third component representing
continuous star formation (SF) for the past 10$^7$ yr at the rate of 
0.16~$M_{\sun}$ yr$^{-1}$ with a normal (Salpeter) IMF, we find that we
can reproduce the M~87 HUT flux.  Although the total accretion rate
from the M~87 cooling flow is estimated at 20--30~$M_{\sun}$ yr$^{-1}$
(White \& Sarazin 1988\markcite{WS88}), the rate of star formation
is generally a small fraction of the cooling rate in giant
ellipticals; the fraction has 
been calculated as high as 15~\% in some cases, but 
$\lesssim$ 1~\% in many others (O'Connell \& McNamara 1988\markcite{OM88}).
In our EHB + PAGB + SF model, 
SEF$_{\rm EHB}$~=~2.8$\times 10^{-2}$ stars yr$^{-1}$ and
SEF$_{\rm PAGB}$~=~0.36 stars yr$^{-1}$.
If we compare these numbers to those in Tables 4 and 5, we see that adding
the young stars means the ratio of EHB to PAGB stars in M~87 
would be more in line with the ratio in M~60.  Although this
small number of young stars would make ($m_{<1000}-m_{1550}$) significantly
bluer, their contribution to the $V$ band would be negligible, since 
this flux is dominated by a much larger number of RGB stars.
Unfortunately, the S/N in the M~87 data is low enough that we cannot put
limits on the contribution of young stars using the \ion{C}{4} and \ion{Si}{4}
absorption.  In contrast, the NGC~1399 data is of higher S/N, and 
the lack of detectable \ion{C}{4} and \ion{Si}{4} absorption in the NGC~1399
spectrum makes star formation an unlikely explanation for its FUV colors
(cf. Ferguson et al.\ 1991\markcite{F91}).

\section{LINE STRENGTHS}

Although the line strengths in our FUV spectra probably do not reflect the
inherent abundances driving the evolution in these galaxies, they can still
tell us information about the abundances in the atmospheres of stars in these
evolved populations.  
We found that EHB + PAGB models with $Z_{atm}$~=~0.1~$Z_{\sun}$
best reproduced the overall absorption features in the HUT data.  
Fig.~8 shows the M~60 data and the
best fit EHB + PAGB model for M~60 (Table 4), but with three different
atmospheric metallicities:
$Z_{atm}$~=~$Z_{\sun}$, 0.1~$Z_{\sun}$, and 0.01~$Z_{\sun}$. 
It is obvious that the practically featureless
0.01~$Z_{\sun}$ model (bottom panel) has absorption
lines that are too weak to match the HUT data.  In contrast,
the $Z_{\sun}$ model (top panel) 
has absorption lines which are mostly too strong.
Most of the absorption lines are best matched in the 0.1~$Z_{\sun}$ model
(center panel), although the broad 
blanketing from Fe absorption at 1550~\AA\ and 1000~\AA\
might be better matched in the $Z_{\sun}$ model.  Thus, in M~60, the
atmospheric metal abundance is likely to be somewhere in the range
0.1--1~$Z_{\sun}$.

To quantify the atmospheric abundances in these populations and to look
for any correlation between the line strengths and colors,
we measure the absorption line strengths empirically, by
absorption line fitting and spectral indices. 
These two methods 
do not rely upon stellar evolutionary models or synthetic spectra.
However, in practice, the low S/N in the HUT data hampers these measurements
and makes it difficult to draw strong conclusions regarding the
atmospheric abundances in these populations.

\subsection{Spectral Indices}

By using the EHB models as a guide, we can define spectral indices which
reflect the abundances of elements responsible for strong absorption features
in the FUV: C, N, Si, and Fe.  While C, N, and Si each have a small number
of strong features in the HUT wavelength range, Fe produces strong line
blanketing over large regions of this range (Brown et al.\
1996\markcite{BFD96}).    

We define line and continuum regions for C, N, Si, and Fe, and then measure
the spectral index $I$ in magnitudes as

\[I = 2.5~log_{10}~ \frac{\displaystyle N_{cont}^{-1} \sum_{cont} F_{\lambda}}{\displaystyle N_{lines}^{-1} \sum_{lines} F_{\lambda}},\]

\noindent
where $N_{cont}$ and $N_{lines}$ are the number of continuum bins and
line bins, respectively.

Table 6 lists the line and continuum regions used to define our spectral 
indices.  Because the FUV is completely blanketed by absorption lines,
there is no way to define a ``clean'' spectral index that will be affected
by absorption from only a single element (see Figs. 9--12 in Brown et 
al.\ 1996\markcite{BFD96}).  To show how these indices may change as 
a function of abundance in a stellar atmosphere, we measured the indices
in an LTE synthetic stellar spectrum with 
T$_{\rm eff}$~=~25,000~K, log~g~=~5.0,
and $Z_{atm}$~=~$Z_{\sun}$.
We repeated this measurement in synthetic spectra where one of four
elements (C, N, Si, and Fe) was missing, and the results are plotted in
Fig.~9.  Because the indices are not completely ``clean,''  the index from
one element can be affected by the abundance of other elements
(see for example the effect of removing Fe in the upper right panel
of Fig.~9).  However,
it is apparent in the figure that these indices are sensitive to the 
individual abundance changes in these four elements.

To improve the S/N, we combined the deredshifted
spectra from galaxies with strong UV upturns (M~60 and M~89) to
produce a ``UV-strong'' spectrum, and 
the deredshifted spectra from galaxies with weak UV upturns (M~49, NGC~3115,
and NGC~3379) to produce a ``UV-weak'' spectrum.  The individual spectra
were weighted by the variance in this process.  We then measured the spectral
indices in both the UV-strong and UV-weak spectra.  The UV-strong and
UV-weak spectra are plotted in Fig.~10, along with the regions used for
the spectral index measurements.

Fig.~11 shows the variation in these indices 
vs. $Z_{atm}$, for each of the EHB + PAGB models shown in Fig.~6
and Table 4.
In this way, we can account for the spread in indices due to the uncertainties
in $Z_{evol}$: each panel shows the indices as a function of 
$Z_{atm}$ for a model from
each of the $Z_{evol}$ groups in Table 2.  On the left-hand
side of each panel, we plot the indices for the UV-weak and UV-strong
Astro-2 data.  These indices are shown as error bars that encompass the
1$\sigma$ statistical uncertainties.
Because these uncertainties are fairly large, we cannot use the indices
to determine accurately what the atmospheric abundances are in the
stellar populations in these galaxies.  However, it is evident in Fig.~11
that the indices are consistent with $Z_{atm} \approx 0.1$~$Z_{\sun}$.
It is also apparent that the HUT data exhibit a weak trend, such that
the UV-strong galaxies tend to have stronger UV spectral indices than
the UV-weak galaxies; seven of the nine indices in Fig.~11 show this
tendency.  

\subsection{Absorption Line Fitting}

We attempted to fit absorption
lines in the two Astro-2 galaxies with reasonable S/N: M~60 and NGC~3379.  
Each of these
galaxies was observed multiple times during the Astro-2 mission, in order
to improve the S/N in two galaxies with very different UV upturn strengths.
We also fit these same features in the best-fit EHB + PAGB model for M~60
($Z_{atm}$~=~0.1~$Z_{\sun}$; Table 4).  
The fits to the two galaxies and the model
are listed in Table~7.

The low S/N of the Astro-2 data required that we restrict the number of
free parameters in our absorption line fits.  Thus, we fit Gaussian 
absorption profiles to those features listed in Table~7, holding the
FWHM and the line center fixed.  The FWHM was held to the quadrature sum
of the intrinsic line width (determined from the EHB + PAGB model), the 
velocity dispersion in the nucleus of each galaxy, and the 3~\AA\ resolution
of the HUT data.  The velocity dispersions of M~60 and NGC~3379 are
363 and 214 km s$^{-1}$ respectively (Burstein et al.\ 1988\markcite{B88}).
The uncertainties in equivalent width (EW) are calculated from the
1$\sigma$ errors, the upper limits in EW are calculated for the 2$\sigma$
confidence level, and the significance is the formal probability that
the absorption feature results from a statistical fluctuation.

An analysis of the line strengths in Table 7 does not provide any indication
of systematic changes between the 
absorption lines in the UV-strong M~60 and those in the UV-weak NGC~3379.
However, given the statistical uncertainty in these measurements, a
correlation between the FUV colors and the FUV absorption line strengths could 
easily be lost in the noise.  The absorption lines are again
consistent with Z$_{atm}\approx$~0.1~$Z_{\sun}$.  

\section{DISCUSSION AND SUMMARY}

Each individual analytical technique we have employed
provides evidence for somewhat weak conclusions about the 
chemical composition of the stellar populations
in these galaxies. Taken in their entirety, our results
favor the hypothesis that
the FUV light from elliptical galaxies originates in a population with
high metallicity and possibly high He abundance.
These conclusions appear to depend on the ratio of post-HB
to EHB flux derived from the tracks.  The results for the tracks with
$Z_{evol} >$~$Z_{\sun}$ 
are less distinguished from each other than they are from those
of lower metallicity in this respect, so that more specific conclusions
cannot be drawn.  We cannot, however, from the FUV spectra alone, 
rule out the low-metallicity scenario of Park \& Lee (1997\markcite{PL97}).

The preceding sections have demonstrated:\\
\noindent
1) The HUT Astro-1 and Astro-2 E galaxy spectra are best fit by a
composite EHB + PAGB star model; the FUV spectra are dominated by
EHB stars following AGBM evolution.\\
\noindent
2) When the flux from the PAGB component is constrained by the fuel
consumption theorem, EHB stars of high $Y_{ZAMS}$ and $Z_{evol}$ are
favored to make up the flux deficit near the Lyman limit.\\
\noindent
3) Absorption features are consistent with $Z_{atm} \simeq$~0.1~$Z_{\sun}$
and may tend to increase as the UV upturn increases,
but the direct constraints on metallicity from spectral lines are weak due
to the low S/N of the data and the likely redistribution of elements
within the EHB star atmospheres.

If EHB stars alone could account for the FUV light in elliptical 
galaxies, then EHB stars with high $Z_{evol}$ and high $Y_{ZAMS}$
would best fit the Astro-2 data.  However, we know that PAGB stars 
exist in ellipticals; we can observe bright PAGB stars in elliptical
galaxies, and the fuel consumption theorem tells us that the stars evolving
from the EHB are only a small fraction of the entire evolved population.
If all of the stars in the population evolved from the
EHB, the FUV light from ellipticals would be much brighter
than that observed, since EHB stars are such efficient FUV emitters.
Dorman et al.\ (1995\markcite{DOR95}) 
estimate $(m_{1550}-V) \sim 0$ for systems where
all of the HB stars lose sufficient mass on the RGB
to become EHB and subsequently AGBM stars (i.e. sdB and then sdO stars).
Hence the theoretical bluest possible ($m_{1550}-V$) color implies 
$\sim 6$ times more FUV radiation than is observed in the nucleus of
bluest galaxy, NGC~1399.  
Since we know PAGB stars make up the bulk of the FUV population and
contribute to at least a fraction of the FUV light, a more realistic
approach is a two-component EHB + PAGB model.  If the EHB and PAGB components
are fit to the HUT data without 
fuel consumption constraints, then the HUT data can be
reproduced equally well by populations with low $Z_{evol}$ \& $Y_{ZAMS}$
or high $Z_{evol}$ \& $Y_{ZAMS}$.  The ambiguity occurs because a ``hot'' PAGB
spectrum can offset the flux deficit at short wavelengths in the
low $Z_{evol}$ and $Y_{ZAMS}$ models.  However, an unconstrained EHB + PAGB 
model is unrealistic, because
in such fits the number of PAGB stars exceeds that allowed by the
fuel consumption theorem.   Although the EHB and 
unconstrained EHB + PAGB models are instructive,
a more realistic model is one where
the number of PAGB stars is astrophysically plausible and thus constrained
by the fuel consumption
theorem.  When we fit these constrained EHB + PAGB models
to the HUT data, the fitting favors a population 
evolving with high $Z_{evol}$ and high $Y_{ZAMS}$.

The EHB stars that best fit the Astro-1 and Astro-2 data in any 
EHB or EHB + PAGB 
model are those that lie within a narrow distribution of envelope mass on the
low-mass (blue) end of the HB.  The stars in this narrow distribution evolve
along AGBM paths,
but the mass distribution does not include those EHB 
stars with the very lowest 
envelope masses on the HB.  In addition,
since the SEDs of PEAGB stars are so
dissimilar to the Astro-2 data (see Fig.~1), PEAGB stars 
cannot comprise a significant
fraction of the stellar populations in these galaxies.
This result 
is partly to be expected from the small mass range that produces them
for a given $Y$ and $Z$.  However, it is possible (and plausible) that
relatively low mass PEAGB stars (with $\log L/L_{\sun}
\sim 3.2)$ are produced 
in old populations through mass loss on the AGB.  Such objects are in
principle capable of supplying the total UV flux for the weaker UV upturn
systems (with $m_{1500}-V \gtrsim 3.6$).  Our modeling implies that
all of the galaxies we have studied instead have some significant 
contribution from a small EHB population.  From this we infer that
the galaxies have a bimodal HB population in general.
Nesci \& Perola (1985\markcite{NP95}) showed that a strongly 
bimodal temperature
distribution on the HB was needed to explain the 2200~\AA\ dip seen in IUE 
spectra of elliptical galaxies, and
Ferguson (1994\markcite{F94}) invoked
a bimodal temperature distribution to explain the SEDs of the
Astro-1 observations of NGC~1399 and M~31, since a uniform distribution of mass
on the HB would produce a ``flatter'' spectrum (Ferguson 1994\markcite{F94})
than observed.  

The open cluster NGC~6791 might be an analog of the situation
in elliptical galaxies, since it shows a 
high-metallicity evolved population with a strongly bimodal temperature
and mass distribution on the HB (Liebert et al.\ 1994\markcite{LSG94}).  
However, the existence of this cluster might present problems to
the theorists in both the high-metallicity and low-metallicity 
camps in the elliptical galaxy composition debate.  
It is also possible that this cluster {\it may not} be providing us 
direct information about mass loss in the single stars of metal
rich systems, since several of the observed stars are binaries
(E. M. Green 1996, private communication) which may have affected
the evolution of the giant before core helium ignition, by 
tidal interaction and by mass ejection.  However, bimodality
occurs in globular cluster HB morphology as well
(e.g. NGC~1851; Walker 1992\markcite{WA92}), 
where binarism is much less common.
We currently
lack the understanding to 
explain the existence of a bimodal mass distribution
in a coeval, single-metallicity population.  The mass-loss mechanisms
on the RGB that determine the distribution of mass on the HB are
the largest source of uncertainty in the evolution of these 
populations.  

Given a bimodal HB distribution where the efficient UV emitters
are on the EHB,
the ($m_{1550}-V$) color should track that fraction of
stars evolving from the EHB, while the (m$_{<1000}-m_{1550}$) color should
track the ``characteristic'' temperature of the FUV population.  
The uncertainties in mass distribution on the HB
means that we will need more than seven
data points to understand the
behavior in a color-color diagram like that shown in Fig.~7.  
However, these FUV colors are consistent with the scenario
proposed by Ferguson \& Davidsen (1993\markcite{FD93}) that 
UV-weak galaxies have a relatively large PAGB contribution to their spectra,
while the UV-strong galaxies are dominated by EHB stars and their AGBM 
progeny, evolving from a range of temperatures on the blue end of the HB.
In the future, with more data points in such a color-color diagram, we 
might be able to deduce information about the HB mass distribution.

Park \& Lee (1997\markcite{PL97}) argue that stars in the 
low-metallicity tail
of an extended metallicity distribution are responsible for the UV upturn
seen in elliptical galaxies.  
Bressan et al.\ (1994\markcite{BCF94})
believe that a high-metallicity population produces the UV flux
in these galaxies, and they explore metallicities as high as
$Z$~=~0.05.  If the absorption features in evolved populations
reflect their intrinsic abundances, then the actual situation in
elliptical galaxies lies somewhere in between these two extremes,
since the absorption features in the HUT data are consistent with 
$Z_{atm} \simeq$~0.1~$Z_{\sun}$.  
There is a weak trend in the spectra of these 
galaxies, such that galaxies with stronger UV upturns show stronger UV
line indices.   This trend is in the opposite sense to that predicted
by Park \& Lee (1997\markcite{PL97}), but in agreement with the predictions
of Bressan et al.\ (1994\markcite{BCF94}) and Dorman 
et al.\ (1995\markcite{DOR95}).  

However, diffusion processes in the envelopes of stars in
evolved populations tend to enhance low intrinsic abundances and
diminish high intrinsic abundances, and thus it is unlikely that
$Z_{atm}$~=~$Z_{evol}$.  Given this situation, we must also compare the 
spectral energy distributions predicted by modeling the FUV data.  This
comparison supports the theory that the strength of the UV upturn
in elliptical galaxies is directly tied to that fraction of a 
high $Z_{evol}$ and high $Y_{ZAMS}$ 
population evolving along AGBM tracks from the
blue end of the horizontal branch.  Integrations over high $Z_{evol}$ and
high $Y_{ZAMS}$ evolutionary tracks best fit the HUT data, and AGBM evolution
is more easily produced in a high $Z_{evol}$ and high Y$_{ZAMS}$ 
population, providing
a natural explanation for the correlation between UV upturn strength
and metallicity found by Burstein et al.\ (1988\markcite{B88}).

Of course, the large uncertainties in our analysis mean that we cannot
rule out the Park \& Lee (1997\markcite{PL97}) hypothesis.  The 
$\chi^2$ statistics in our fitting favor the high $Y$ \& $Z_{evol}$ models, but
to the eye one can produce reasonable matches to the HUT data
with integrations over low $Y$ \& $Z_{evol}$ tracks.  In the Park \& Lee work,
they argue that the FUV flux is dominated by a population with 
[Fe/H]~$<$~--0.7, but unfortunately 
it is difficult to discern from their paper the exact
distributions in $M_{env}$ and metallicity which are used in their models.
Furthermore, although Park \& Lee stress that their purpose is not to
match the actual SEDs observed in ellipticals, it seems that their synthetic
spectra are too flat in the FUV, at least 
in comparison to the overplotted IUE data
(cf.\ their Figs. 6 \& 10).  It would be interesting to apply their models
to the HUT data, in order to determine if their low metallicity models 
can reproduce the actual SEDs
of observed ellipticals, including the flux near the Lyman limit, since 
our analysis indicates
that the low $Y$ \& $Z$ tracks do not reproduce the short wavelength flux
needed to reproduce the HUT SEDs.  

In order to
answer definitively the debate and disentangle the effects of age
and metallicity on the HB mass distribution, 
data with higher S/N for a larger sample of galaxies is required.  
At higher S/N, it should be obvious if 
diffusion processes are playing a large role in the strengths of the
FUV absorption lines, since observations of evolved stars show lines that
deviate strongly from solar abundances or even scaled solar abundances.
However, we should not restrict our study to the strengths of metallic
lines.  The hydrogen lines are also very sensitive to the effective temperature
and surface gravity of the different HB evolutionary phases.  Inspection
of Fig.~1 shows that PAGB spectra have a much shallower Lyman series than
that seen in AGBM spectra.  Although the S/N in the HUT data was not
sufficient to use the Lyman series as a stellar population probe,
accurate measurements of the flux below
Lyman $\alpha$ (and thus beyond the detection capabilities of HST) will
yield important information about the relative numbers of PAGB and AGBM
stars in a given population.  
Such observations could be made with another flight of HUT or a similar
successor.
Fortunately, future FUV measurements of
elliptical galaxies will be accompanied by the rapid
progress currently seen in the calculation
of synthetic stellar spectra.  
The near future should bring the emergence of fully line-blanketed,
non-LTE model atmospheres and synthetic spectra, and with these models,
another source of uncertainty will be removed from our understanding
of the FUV data.

Once we know the chemical 
composition of the FUV sources in elliptical galaxies,
the resolution of the debate
will have impact beyond the study of stellar evolution.
Elliptical galaxies show great promise as standard candles
and tracers of cosmic evolution.
The understanding of the FUV flux from galaxies will also enable
us to decode the information stored in the longer UV wavelengths
(Dorman \& O'Connell 1996 \markcite{DO96}).
The different composition hypotheses 
each argue for different ages of the Universe and different scenarios for the 
construction of the galaxies. 
Under the Park \& Lee (1997\markcite{PL97})
low-metallicity scenario, the more massive galaxies formed first, in
contrast to the ``bottom-up'' cosmologies wherein the
giant elliptical galaxies formed from the merger of smaller galaxies.  
In this way,
the metallicity debate has consequences that extend to important
issues in cosmology and galaxy evolution.

\acknowledgments

This work was supported by NASA contract NAS 5-27000 to the Johns Hopkins
University.  B. D. acknowledges partial support from NASA RTOP 188-41-51-03.
We wish to thank Jeffrey Kruk  and Gerard Kriss for their assistance with the
reduction of the HUT data.  We also wish to thank 
T. Stecher and the UIT team for providing
the FUV image of M~60, and R. Ciardullo for useful discussion.

The optical M~60 DSS image is based on photographic data of the
National Geographic Society -- Palomar Observatory Sky Survey (NGS-POSS) 
obtained using the Oschin Telescope on Palomar Mountain.  The NGS-POSS was 
funded by a grant from the National Geographic Society to the California 
Institute of Technology.  The plates were processed into the present 
compressed digital form with their permission.  The DSS was 
produced at the Space Telescope Science Institute under US Government grant 
NAGW-2166.

\clearpage

\begin{figure}
\plotone{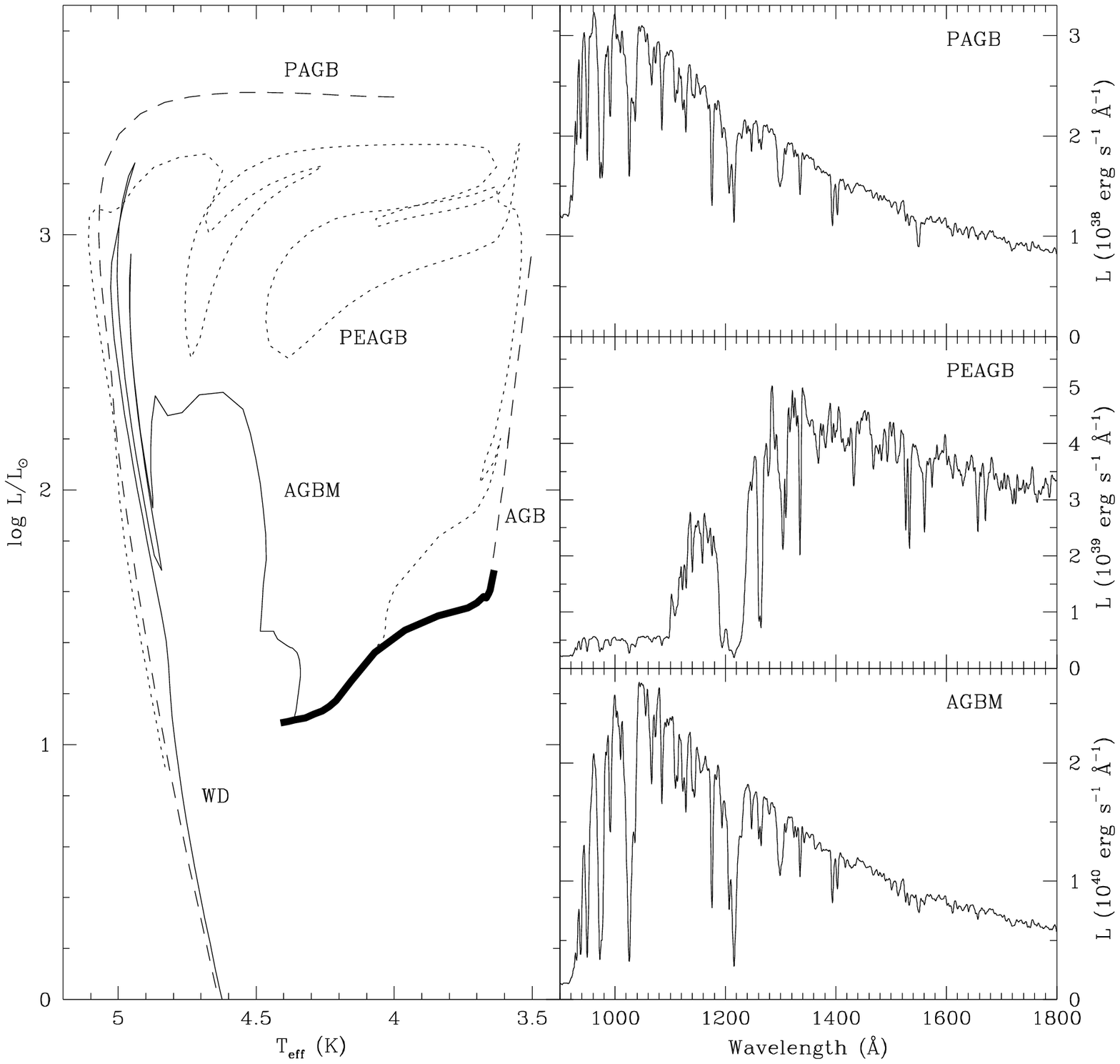}
\caption
{The three classes of evolution for stars on the HB produce
very different FUV spectra when one integrates the FUV luminosity over
these tracks, assuming that stars enter each track at the rate of one
star per year.  The gap between the AGB evolution 
($M_{env}$~=~0.231~$M_{\sun}$; dashed; DRO93) 
and the PAGB evolution 
($M_{core}$=~0.569~$M_{\sun}$; dashed; Vassiliadis \& Wood 
1994) is due to the uncertainties in the TP phase.
All of the evolutionary paths assume solar abundances, and the luminosities 
are integrated over the synthetic spectra grids of 
Brown, Ferguson, \& Davidsen (1996) and Kurucz
(1992), again assuming solar abundances.
Note that the integrated luminosity from the AGBM track
($M_{env}$~=~0.006~$M_{\sun}$; thin solid; DRO93) is much higher than that from
the PEAGB track ($M_{env}$~=~0.051~$M_{\sun}$; dotted; DRO93), 
which in turn is higher than that from the PAGB track (dashed).
The heavy solid line denotes the ZAHB.
}
\end{figure}

\begin{figure}
\plotone{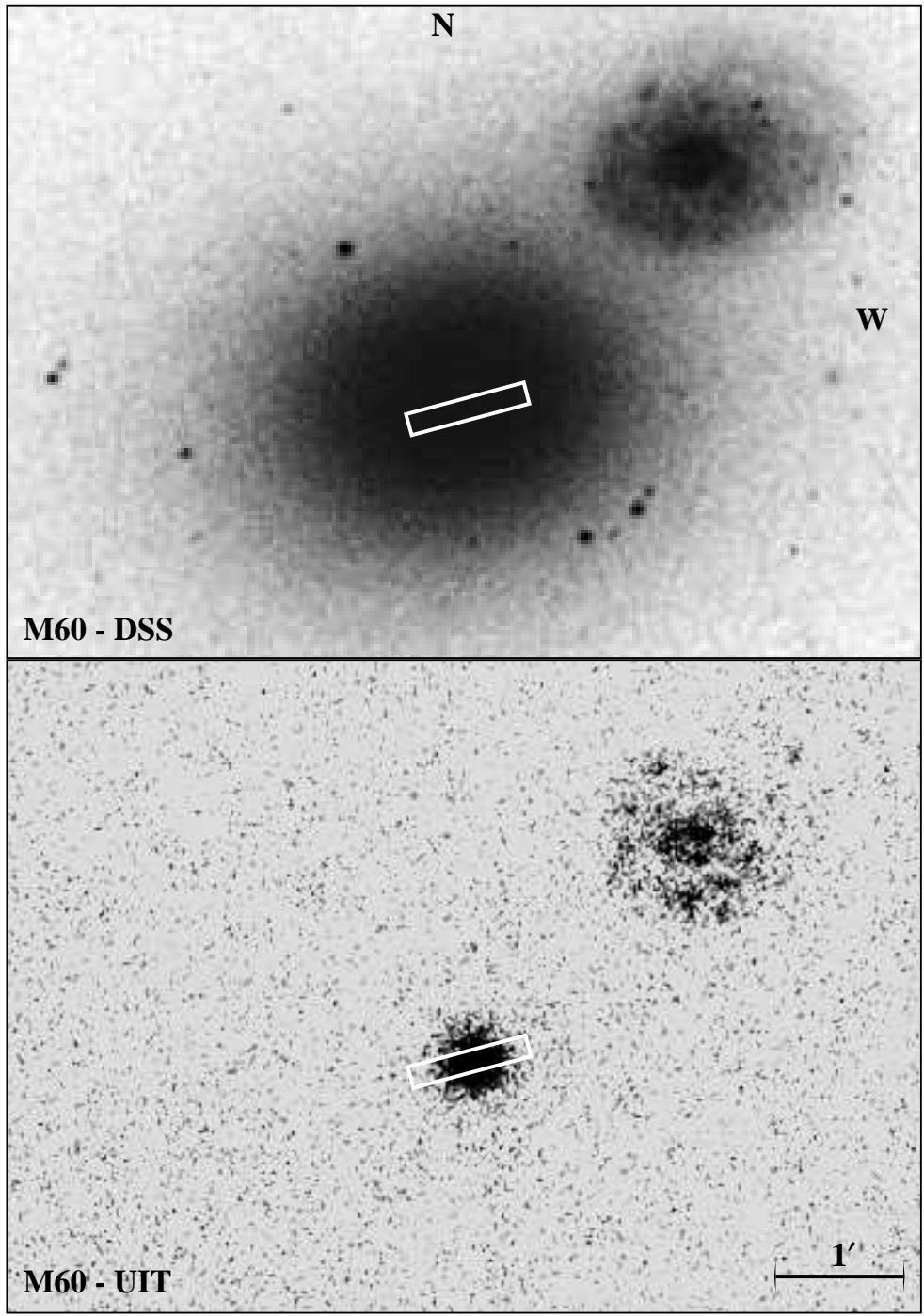}
\caption{
The optical image (top panel) and the FUV image (bottom panel)
of M~60 and NGC~4647 demonstrate that there is no evidence for tidal
interaction between the two galaxies, that much of the concentrated
FUV light from M~60 fell within the HUT slit (white rectangle),
and that light from NGC~4647 is unlikely to contaminate the HUT data for M~60.
}
\end{figure}

\begin{figure}
\plotone{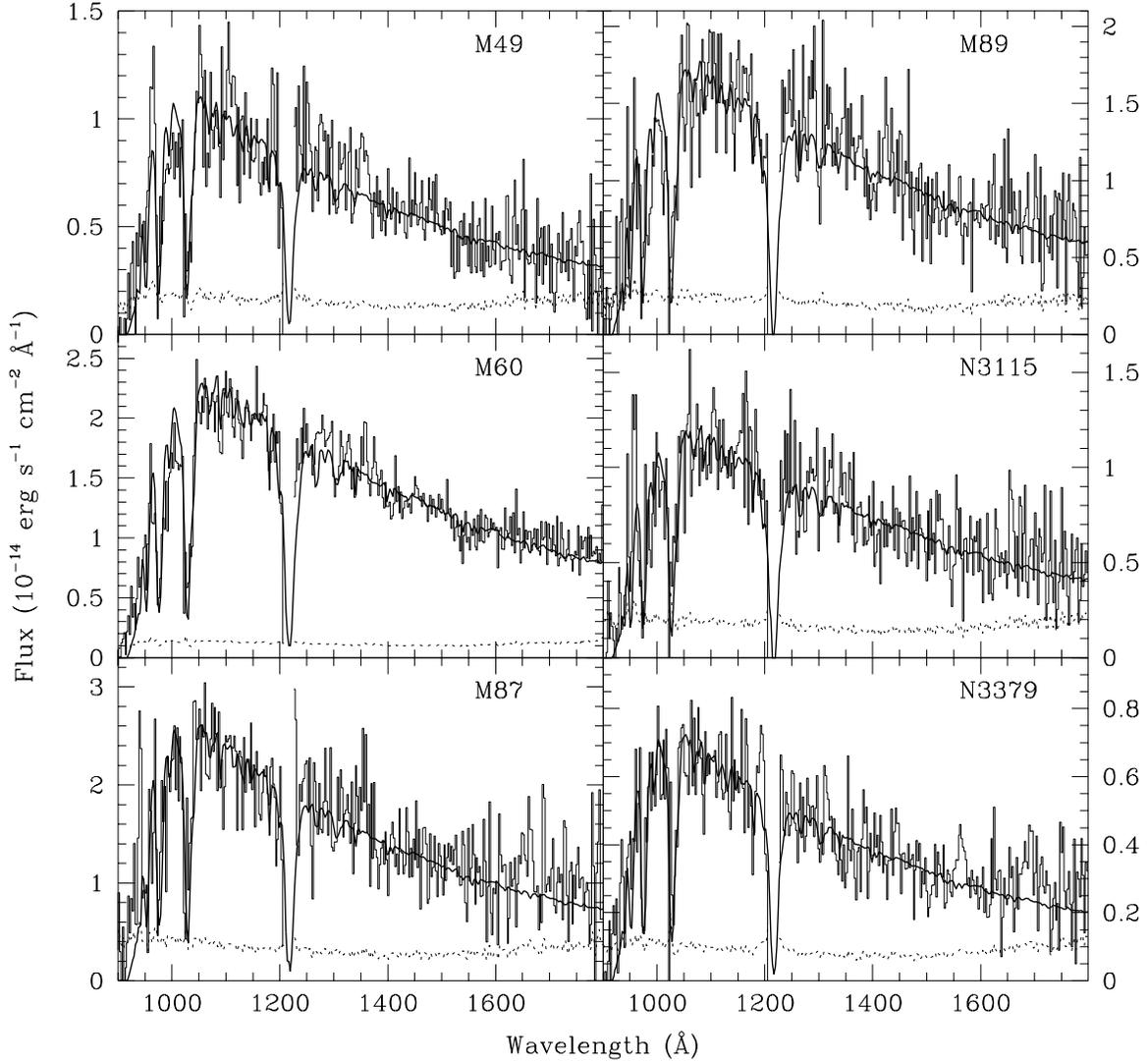}
\caption{
Fig.~3-- The best fit evolutionary model (solid) from integrations over
EHB tracks is plotted over the HUT data (solid; histogram; binned by 2.5~\AA) 
for each of the
Astro-2 observations.  The evolutionary models integrate over synthetic
spectra with $Z_{atm}$~=~0.1~$Z_{\sun}$, which best reproduces the
absorption features in the Astro-2 data, especially in the higher S/N data
of M~60.  Statistical errors in the data are shown dotted.
}
\end{figure}

\begin{figure}
\plotone{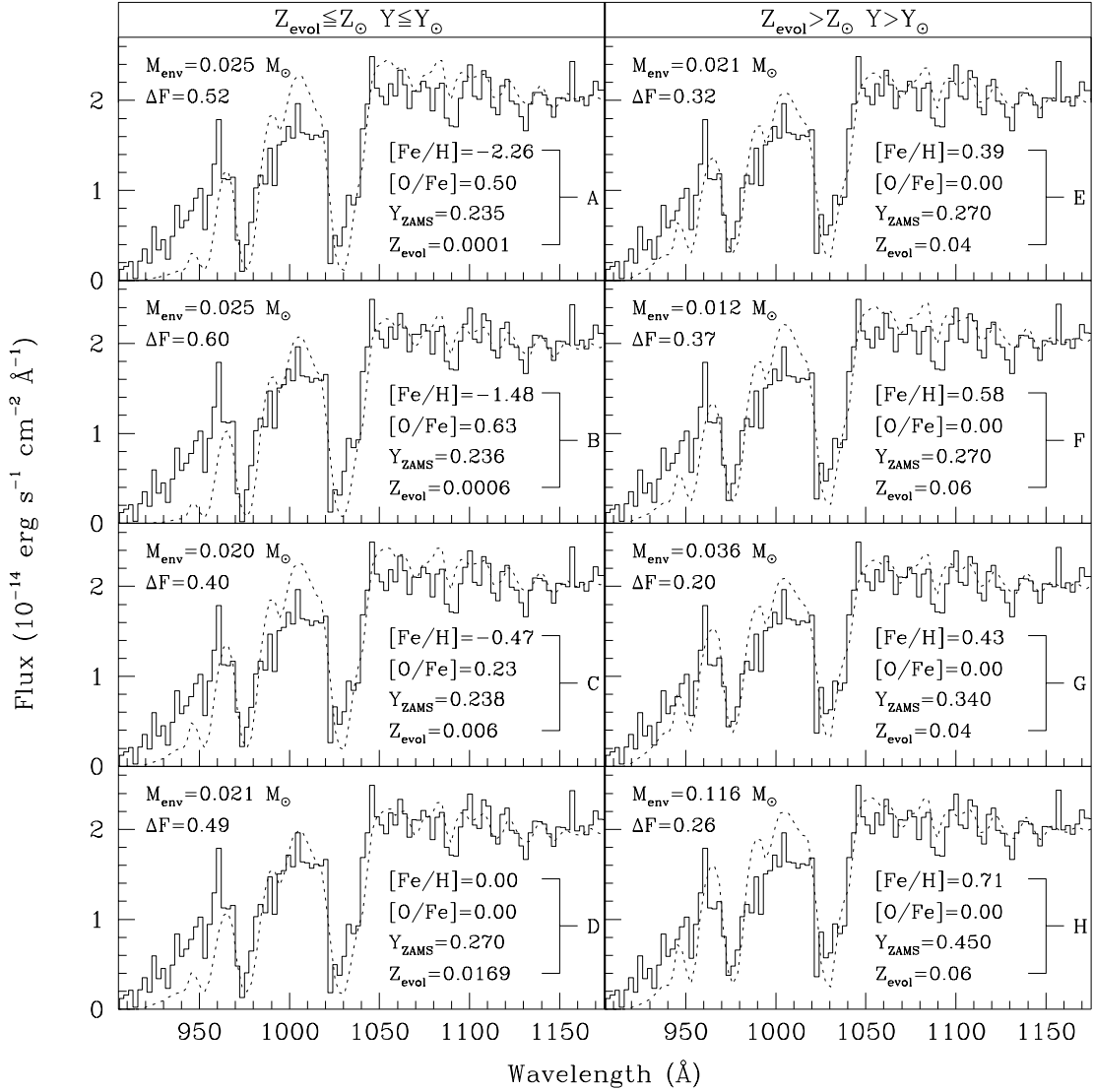}
\caption{
The best fit evolutionary model (dashed) from each of the 
DRO93 abundance subsets (Table 2) 
shows that EHB
stars with high $Z_{evol}$ and high $Y_{ZAMS}$ fit the M~60 data 
(solid histogram)
better than the EHB stars with low $Z_{evol}$ and low $Y_{ZAMS}$.
The abundances and envelope mass driving the evolution 
for each of the evolutionary tracks is
given in each panel, and $Z_{atm}$~=~0.1~$Z_{\sun}$.
Part of the reason the models with supersolar evolutionary abundances 
(right column) fit better than those with subsolar abundances (left column)
is the flux deficiency shortward of 970~\AA\ ($\Delta$F).
}
\end{figure}

\begin{figure}
\plotone{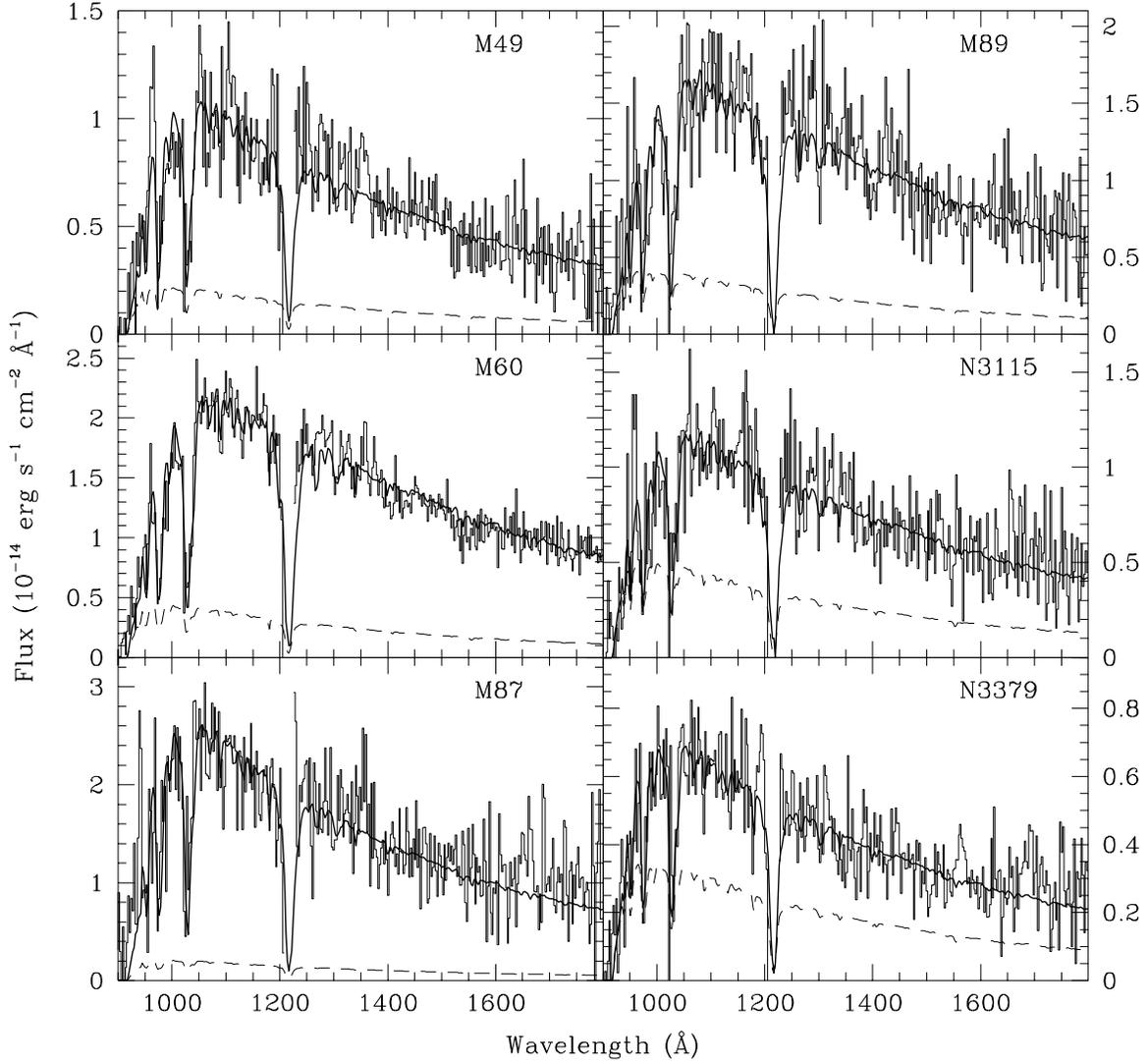}
\caption{
The best fit evolutionary model (solid) from two-component models
integrating over EHB and PAGB tracks is plotted over the HUT data 
(solid histogram; binned by 2.5~\AA) for each of the Astro-2 observations.
As in Fig.~3, the evolutionary models have $Z_{atm}$~=~0.1~Z${\sun}$.
Instead of statistical errors, the PAGB contribution to each two-component
model is shown dashed.
}
\end{figure}

\begin{figure}
\plotone{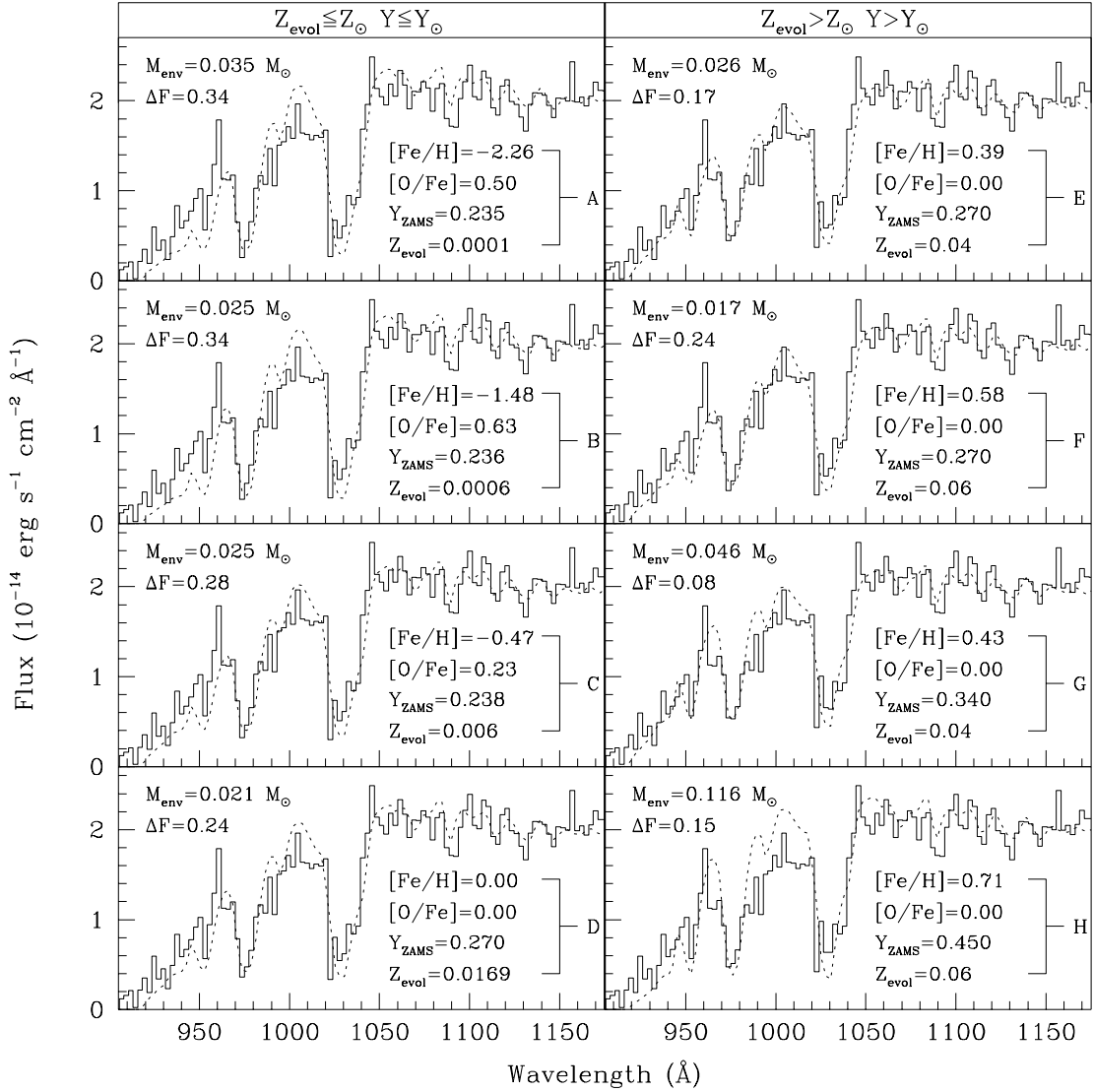}
\caption{
The same as in Fig.~4, except that the model (dashed) 
now has two components: EHB stars and PAGB stars.  The PAGB component
comes from the $Z_{evol}$~=~$Z_{\sun}$ set of Vassiliadis and Wood 
(1994), with $M_{core}$~=~0.569~M${\sun}$.  The legend
on each of the eight plots reflects the abundances which drive the evolution
of the DRO93 EHB tracks.
The models with subsolar evolutionary abundances (left column) again show
significant flux deficiencies shortward of 970~\AA\ ($\Delta$F).
}
\end{figure}

\begin{figure}
\plotone{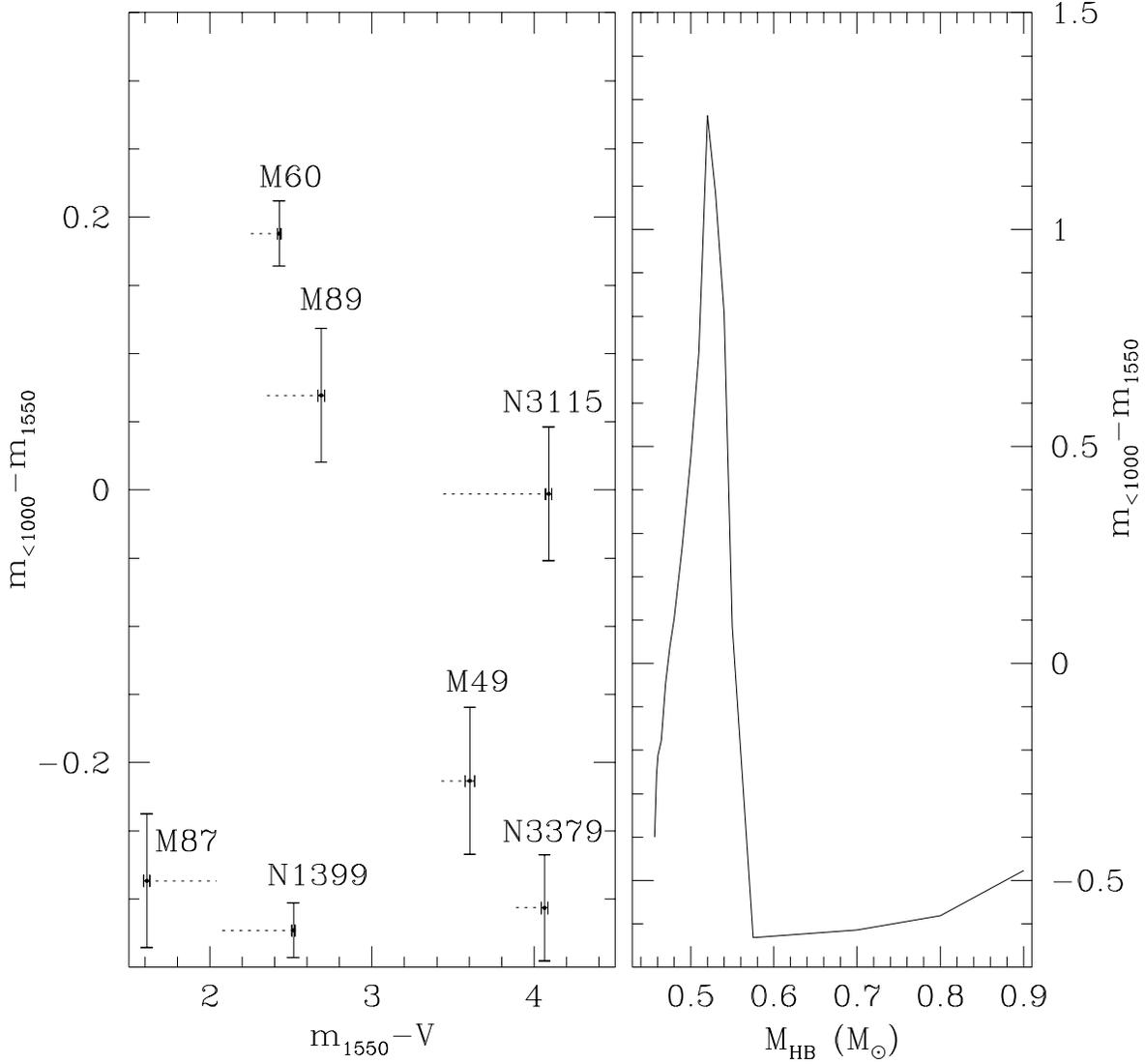}
\caption{
The ($m_{<1000}-m_{1550}$) colors in the integrated synthetic
spectra change considerably
as a function of envelope mass on the HB (right panel).  PAGB stars
evolving from the red end of the HB 
and EHB stars evolving from the extreme blue end of the HB both show
hotter ($m_{<1000}-m{1550}$) colors, 
while stars evolving between these extremes appear cooler.
Thus, in the color-color diagram (left panel), galaxies with
strong UV-upturns (NGC~1399, M~60, M~89)
may be dominated by EHB stars evolving from narrow
ranges of envelope mass on the blue end of the HB, while galaxies
with weaker UV-upturns (M~49, NGC~3115, NGC~3379) may have significant
FUV contributions from PAGB stars.
Error bars show the 1$\sigma$ statistical uncertainty, and the dashed
lines show the shift in ($m_{1550}-V$) color that would occur if
we used the Burstein et al.\ (1988) measurements.
}
\end{figure}

\begin{figure}
\plotone{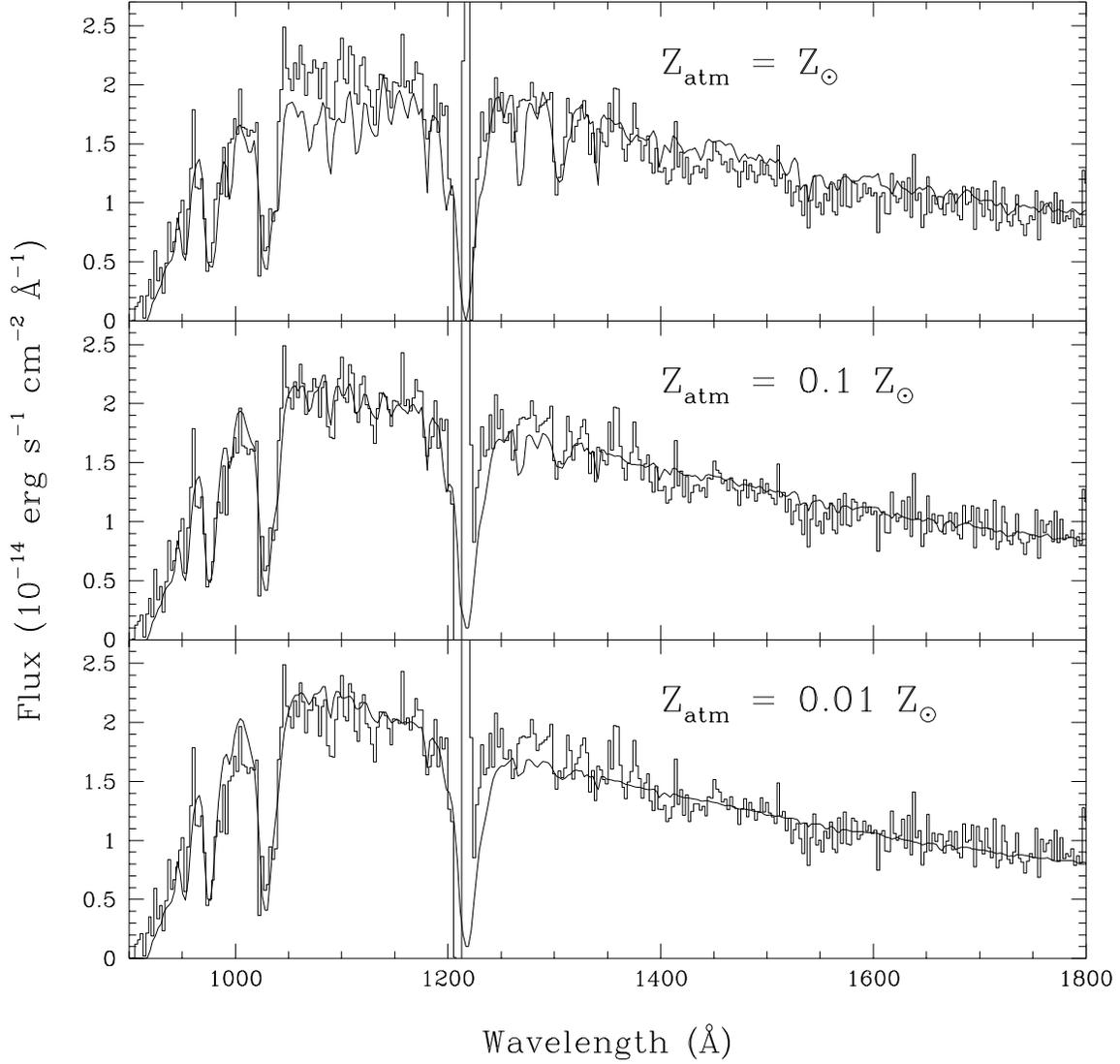}
\caption{
The HUT data for M~60 (histogram) are plotted with the best fit
EHB + PAGB model (solid) for M~60, given in Table 3.  The EHB + PAGB model
integrates over synthetic spectra with 
$Z_{atm}$~=~$Z_{\sun}$, 0.1~$Z_{\sun}$, and 0.01~$Z_{\sun}$.  The absorption
features fit best in the center panel, with $Z_{atm}$~=~0.1~$Z_{\sun}$.
Small differences in the data shown in each panel 
are due to variations in the airglow subtraction,
since airglow profiles were fit simultaneously with the models.
}
\end{figure}

\begin{figure}
\plotone{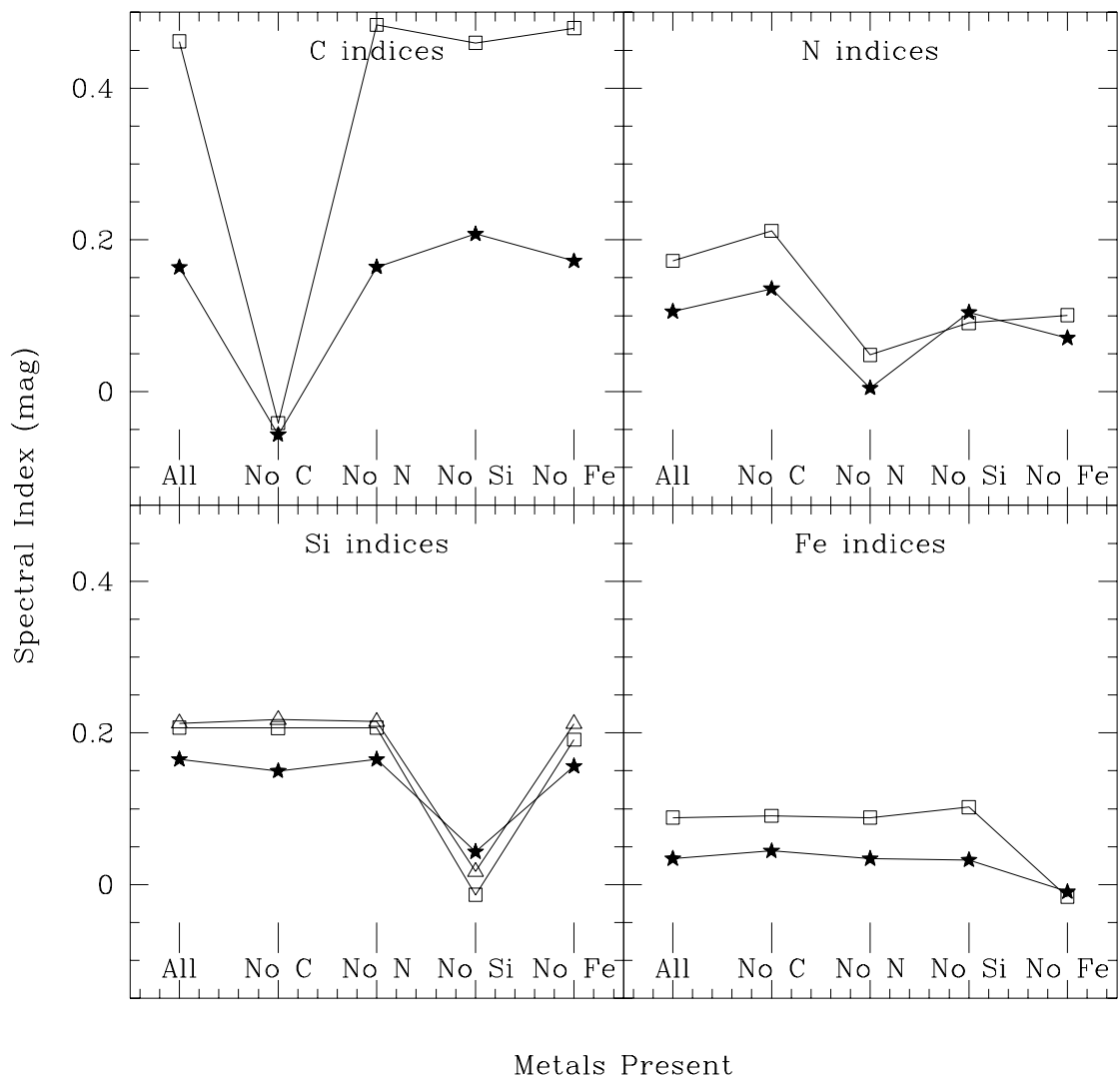}
\caption{
The spectral indices for synthetic spectra with 
T$_{\rm eff}$~=~25,000~K
and log~g~=~5.0 are plotted for spectra at solar abundance
(labeled All) or at solar abundance with one element missing (labeled).
The line blanketing in the FUV prevents the definition of a ``clean''
spectral index that is affected by only one element, but these indices
show the most sensitivity to the intended elements.  The indices are defined
in Table 2.
}
\end{figure}

\begin{figure}
\plotone{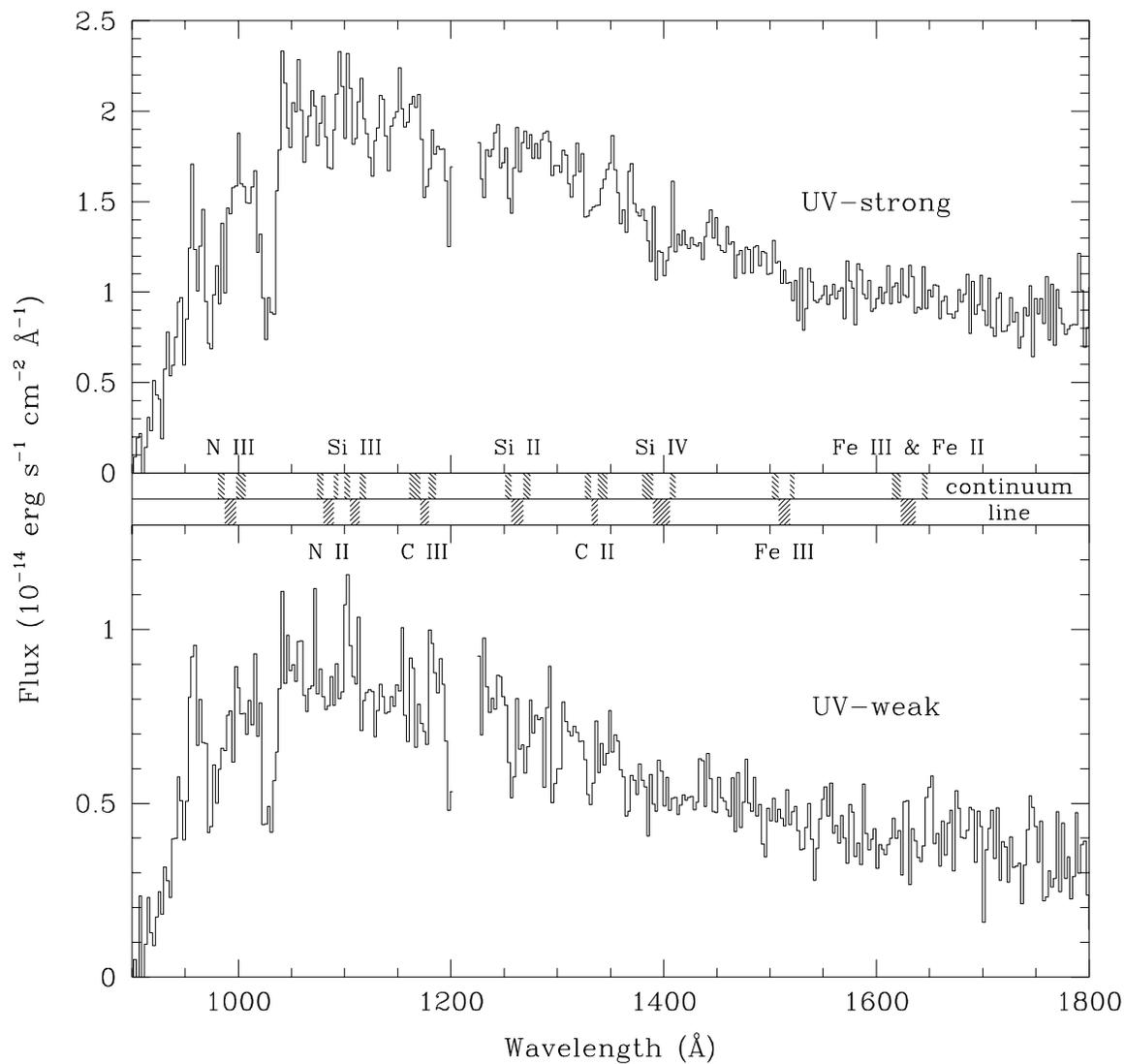}
\caption{
The ``UV-strong'' spectrum (top panel) 
is the variance-weighted average of M~60
and M~89, and the ``UV-weak'' spectrum (bottom panel) is the variance-weighted
average of M~49, NGC~3115, and NGC~3379.  The line and continuum regions
used in the spectral index measurements are indicated.
}
\end{figure}

\begin{figure}
\plotone{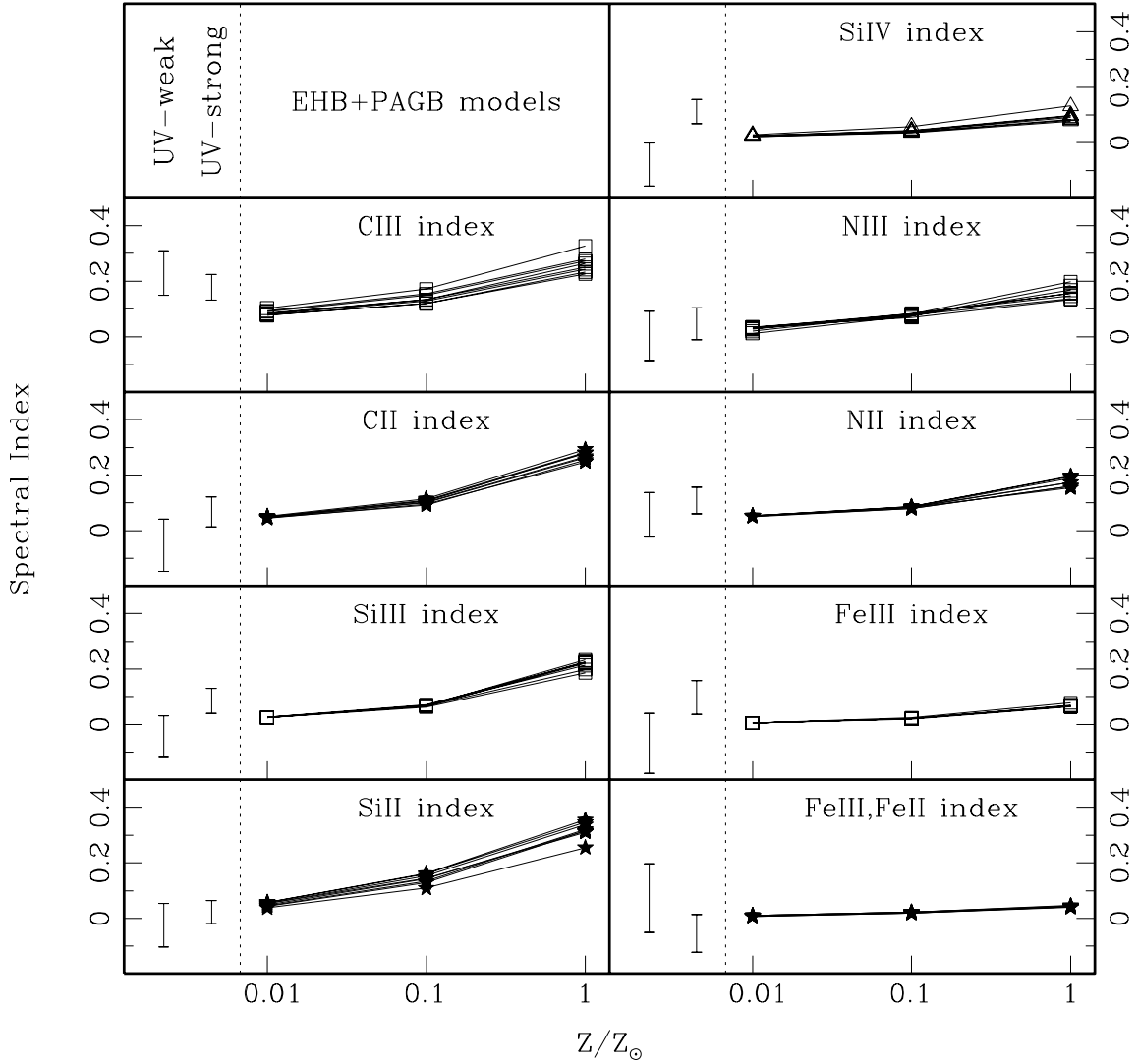}
\caption{
The spectral indices from Table 4, plotted for EHB+PAGB models
and the Astro-2 data.  The UV-weak galaxies (M~49, NGC~3115, and NGC~3379)
were averaged to form a composite UV-weak spectrum, and the UV-strong galaxies
(M~60 and M~89) were averaged to form a composite UV-strong spectrum.
In order to account for spectral index variations due to evolutionary 
assumptions, we have plotted in each panel a given spectral index 
vs. $Z_{atm}$ for eight EHB+PAGB models (one from each of the eight $Z_{evol}$ subsets in Table 2).  
Specifically, each EHB+PAGB model is taken from the best-fit 
models to M~60 that are shown in Fig.~6.  
Note that the data are consistent with low atmospheric abundances ($Z_{atm}$),
with the UV-strong galaxies tending toward stronger indices.
}
\end{figure}

\clearpage 
\begin{table}
{\scriptsize
\caption{HUT Astro-2 Observations}
\begin{tabular}{lcccccc}
\tableline \tableline
& M 49 & M 60 & M 87 & M 89 & NGC 3115 & NGC 3379 \\
\tableline
Exp. (sec) & 1346 & 5224 & 950 & 1682 & 1634 & 3074\\
$D$ (Mpc) & 21.9\tablenotemark{a} & 21.9\tablenotemark{a} & 21.9\tablenotemark{a} & 21.9\tablenotemark{a} & 10.0\tablenotemark{b} & 18.2\tablenotemark{a}\\
$V$ (mag)\tablenotemark{c} & 11.19 & 11.46 & 12.30 & 11.60 & 10.54 & 11.24\\
H I (10$^{20}$ cm$^2$)\tablenotemark{d} & 1.64 & 2.41 & 2.51 & 2.50 & 4.44 & 2.86\\
$E(B-V)$ (mag)\tablenotemark{e} & 0.000 & 0.010 & 0.023 & 0.035 & 0.025 & 0.013\\
($m_{1550}-V$) (mag)\tablenotemark{f} & $3.60\pm0.03$ & $2.43\pm0.01$ & $1.61\pm0.02$ & $2.69\pm0.02$ & $4.09\pm0.02$ & $4.07\pm0.02$ \\
($m_{<1000}-m_{1550}$) (mag)\tablenotemark{g} & $-0.21\pm0.05$ & $0.19\pm0.02$ & $-0.29\pm0.05$ & $0.07\pm0.05$ & $-0.00\pm0.05$ & $-0.31\pm0.04$ \\
\tableline
\tablenotetext{a}{Ferguson \& Sandage 1990\markcite{FS90}}
\tablenotetext{b}{Hanes \& Harris 1986\markcite{HH86}}
\tablenotetext{c}{Through the HUT slit, from archival HST data.}
\tablenotetext{d}{Einstein On-Line Service, Smithsonian Astrophysical Observatory}
\tablenotetext{e}{Burstein \& Heiles 1984\markcite{BH84}}
\tablenotetext{f}{Through the HUT slit, from HUT and archival HST data.}
\tablenotetext{g}{Through the HUT slit.}
\end{tabular}
}
\end{table}

\begin{table}
\caption{Evolutionary Models}
\begin{tabular}{ccccccc}
\tableline \tableline
Set & [Fe/H] & [O/Fe] & $Z_{evol}$ & $Y_{ZAMS}$ & Y$_{HB}$ & M$_{core}$ \\
\tableline
A   & -2.26  & 0.50   & 0.0001     & 0.235      & 0.245    & 0.495   \\
B   & -1.48  & 0.63   & 0.0006     & 0.236      & 0.247    & 0.485   \\
C   & -0.47  & 0.23   & 0.0060     & 0.238      & 0.257    & 0.475   \\
D   &  0.00  & 0.00   & 0.0169     & 0.270      & 0.288    & 0.469   \\
E   &  0.39  & 0.00   & 0.0400     & 0.270      & 0.292    & 0.464   \\
F   &  0.58  & 0.00   & 0.0600     & 0.270      & 0.289    & 0.458   \\
G   &  0.43  & 0.00   & 0.0400     & 0.340      & 0.356    & 0.454   \\
H   &  0.71  & 0.00   & 0.0600     & 0.450      & 0.459    & 0.434   \\
\tableline
\end{tabular}
\end{table}

\begin{table}
{\scriptsize
\caption{Stellar Population Modelling-- EHB Fitting}
\begin{tabular}{lccccccc}
\tableline \tableline
& M 49 & M 60 & M 87 & M 89 & NGC 3115 & NGC 3379 & NGC 1399\\
\tableline
SEF$_{\rm tot}$ (stars yr$^{-1}$)\tablenotemark{a} & 0.94 & 0.76 & 0.36 & 0.72 & 0.39 & 0.65 & 0.84 \\
\tableline
Best $Z_{evol}$ group & H & G & H & G & G & H & H \\
$M_{env}$ ($M_{\sun}$) & 0.086 & 0.036 & 0.076 & 0.036 & 0.036 & 0.076 & 0.016 \\
$\chi^2/\nu$  & 1.2 & 2.9 & 1.7 & 1.7 & 1.3 & 1.3 & 3.0 \\
SEF$_{\rm EHB}$ (stars yr$^{-1}$)\tablenotemark{b}  & 2.1$\times10^{-2}$ & 5.7$\times10^{-2}$ & 5.1$\times10^{-2}$ & 4.3$\times10^{-2}$ & 6.2$\times10^{-3}$ & 9.7$\times10^{-3}$ & 8.1$\times10^{-2}$ \\
$\Delta \chi^2$ for alternative & & & & & & & \\
abundance groups\tablenotemark{c}: & & & & & & & \\
A & 39 & 371 & 64 & 47 & 43 & 47 & 211 \\
B & 30 & 370 & 51 & 50 & 40 & 37 & 142 \\
C & 20 & 218 & 40 & 25 & 26 & 35 &  97 \\
D & 21 & 181 & 40 & 27 & 29 & 35 &  81 \\
E & 17 &  34 & 34 &  3 &  9 & 31 &  71 \\
F & 24 & 176 &  6 & 18 & 21 & 18 &  65 \\
G & 4  &   0 & 11 &  0 &  0 & 15 &  40 \\
H & 0  &  88 &  0 & 13 & 10 &  0 &   0 \\
\tableline
\tablenotetext{a}{Through the HUT slit, based on the fuel consumption
theorem and $V$-magnitude from Table 1.}
\tablenotetext{b}{Through the HUT slit, based upon fitting the HUT FUV
data.}
\tablenotetext{c}{By statistical comparison to the best-fit EHB model
for each galaxy.
$\nu$ is 272 for NGC~1399, and 326 for the Astro-2 galaxies.}
\end{tabular}
}
\end{table}

\begin{table}
{\scriptsize
\caption{Stellar Population Modelling-- EHB + PAGB Fitting (M$_{core}^{PAGB}$~=~0.569~$M_{\sun}$)}
\begin{tabular}{lccccccc}
\tableline \tableline
& M 49 & M 60 & M 87 & M 89 & NGC 3115 & NGC 3379 & NGC 1399\\
\tableline
SEF$_{\rm tot}$ (stars yr$^{-1}$)\tablenotemark{a} & 0.94 & 0.76 & 0.36 & 0.72 & 0.39 & 0.65 & 0.84 \\
\tableline
Best $Z_{evol}$ group & G\&H & E & H & D\&F & B\&F & C\&H & H \\
$M_{env}$ ($M_{\sun}$) & 0.026 & 0.026 & 0.056 & 0.017 & 0.017 & 0.116 & 0.016\\
$\chi^2/\nu$  & 1.2 & 2.6 & 1.6 & 1.7 & 1.3 & 1.2 & 3.0\\
M$_{env}$ 4$\sigma$ limits\tablenotemark{b} & 0.007--0.116 & 0.021--0.046 & 0.016--0.106 & 0.012--0.046 & 0.012--0.166 & 0.007--0.166 & 0.011--0.026 \\
SEF$_{\rm EHB}$ (stars yr$^{-1}$)\tablenotemark{c}  & 2.1$\times10^{-2}$ & 5.6$\times10^{-2}$ & 3.9$\times10^{-2}$ & 4.1$\times10^{-2}$ & 4.7$\times10^{-3}$ & 4.8$\times10^{-3}$ & 6.8$\times10^{-2}$ \\
SEF$_{\rm PAGB}$ (stars yr$^{-1}$)\tablenotemark{c} & 0.38 & 0.76 & 0.36 & 0.72 & 0.18 & 0.41 & 0.84\\
$\Delta \chi^2$ for alternative & & & & & & & \\
abundance groups\tablenotemark{d}: & & & & & & & \\
A & 5 & 131 & 47 & 13 & 3 & 1 & 211 \\
B & 3 & 118 & 43 & 15 & 0 & 1 & 142 \\
C & 1 &  37 & 30 &  4 & 2 & 0 &  97 \\
D & 2 &  16 & 32 &  0 & 3 & 1 &  81 \\
E & 2 &   0 & 24 &  2 & 2 & 3 &  71 \\
F & 2 &  21 &  7 &  0 & 0 & 2 &  65 \\
G & 0 &   3 &  6 &  8 & 5 & 4 &  40 \\
H & 0 & 144 &  0 & 27 & 7 & 0 &   0 \\
\tableline
\tablenotetext{a}{Through the HUT slit, based on the fuel consumption
theorem and $V$-magnitude from Table 1.}
\tablenotetext{b}{In any $Z_{evol}$ group.}
\tablenotetext{c}{Through the HUT slit, based upon fitting the HUT FUV
data.}
\tablenotetext{d}{By statistical comparison to the best-fit EHB+PAGB model
for each galaxy.
$\nu$ is 271 for NGC~1399, and 325 for the Astro-2 galaxies.}
\end{tabular}
}
\end{table}

\begin{table}
{\scriptsize
\caption{Stellar Population Modelling-- EHB + PAGB Fitting (M$_{core}^{PAGB}$~=~0.597~$M_{\sun}$)}
\begin{tabular}{lccccccc}
\tableline \tableline
& M 49 & M 60 & M 87 & M 89 & NGC 3115 & NGC 3379 & NGC 1399\\
\tableline
SEF$_{\rm tot}$ (stars yr$^{-1}$)\tablenotemark{a} & 0.94 & 0.76 & 0.36 & 0.72 & 0.39 & 0.65 & 0.84 \\
\tableline
Best $Z_{evol}$ group & G & G & H & E & F & H & H \\
$M_{env}$ ($M_{\sun}$) & 0.026 & 0.046 & 0.056 & 0.021 & 0.017 & 0.076 & 0.021\\
$\chi^2/\nu$  & 1.2 & 2.7 & 1.6 & 1.7 & 1.3 & 1.2 & 3.0\\
M$_{env}$ 4$\sigma$ limits\tablenotemark{b} & 0.011--0.106 & 0.046--0.046 & 0.016--0.106 & 0.012--0.116 & 0.017--0.166 & 0.011--0.86 & 0.011--0.026 \\
SEF$_{\rm EHB}$ (stars yr$^{-1}$)\tablenotemark{c}  & 2.1$\times10^{-2}$ & 5.3$\times10^{-2}$ & 5.1$\times10^{-2}$ & 4.1$\times10^{-2}$ & 4.9$\times10^{-3}$ & 7.5$\times10^{-3}$ & 9.7$\times10^{-2}$ \\
SEF$_{\rm PAGB}$ (stars yr$^{-1}$)\tablenotemark{c} & 0.018 & 0.76 & 0.36 & 0.71 & 0.33 & 0.38 & 0.84\\
$\Delta \chi^2$ for alternative & & & & & & & \\
abundance groups\tablenotemark{d}: & & & & & & & \\
A & 30 & 233 & 55 & 22 & 7 & 20 & 211 \\
B & 26 & 217 & 47 & 23 & 5 & 17 & 142 \\
C & 10 & 130 & 35 & 13 & 3 &  5 &  97 \\
D & 15 &  72 & 36 &  3 & 4 & 40 &  81 \\
E &  3 &  39 & 28 &  0 & 1 & 12 &  71 \\
F & 20 & 107 &  4 &  9 & 0 & 24 &  65 \\
G &  0 &   0 &  8 &  3 & 3 &  1 &  40 \\
H &  3 & 123 &  0 & 16 & 5 &  0 &   0 \\
\tableline
\tablenotetext{a}{Through the HUT slit, based on the fuel consumption
theorem and $V$-magnitude from Table 1.}
\tablenotetext{b}{In any $Z_{evol}$ group.}
\tablenotetext{c}{Through the HUT slit, based upon fitting the HUT FUV
data.}
\tablenotetext{d}{By statistical comparison to the best-fit EHB+PAGB model
for each galaxy.
$\nu$ is 271 for NGC~1399, and 325 for the Astro-2 galaxies.}
\end{tabular}
}
\end{table}

\begin{table}
\caption{Spectral Indices}
\begin{tabular}{lccc}
\tableline \tableline
Ion & Line region (\AA) & Continuum regions (\AA) & Plot symbol\tablenotemark{a} \\
\tableline
\ion{C}{3} & 1171--1179 & 1161--1171, 1179--1186 & $\Box$ \\
\ion{C}{2} & 1332--1338 & 1326--1332, 1338--1347 & $\star$ \\
\ion{N}{3} & 987--998   & 981--987, 998--1007    & $\Box$ \\
\ion{N}{2} & 1080--1090 & 1074--1080, 1090--1094 & $\star$ \\
\ion{Si}{3} & 1105--1114 & 1100--1105, 1114--1120 & $\Box$ \\
\ion{Si}{2} & 1257--1268 & 1251--1257, 1268--1275 & $\star$ \\
\ion{Si}{4} & 1390--1406 & 1380--1390, 1406--1411 & $\triangle$ \\
\ion{Fe}{3} & 1508--1519 & 1502--1508, 1519--1523 & $\Box$ \\
\ion{Fe}{3}, \ion{Fe}{2} & 1623--1637 & 1615--1623, 1643--1648 & $\star$ \\
\tableline
\tablenotetext{a}{In Figs. 9 and 11.}
\end{tabular}
\end{table}

\begin{table}
\caption{Absorption Lines}
\begin{tabular}{lcccccc}
\tableline \tableline
& &\multicolumn{4}{c} {M 60 \hskip 8em NGC 3379} & Model \\
Feature & $\lambda_o$ & EW(\AA)\tablenotemark{a} & Significance & EW (\AA)\tablenotemark{a} & Significance & EW(\AA)\tablenotemark{b} \\ 
\tableline

Ly$\delta$ &  949.8 & $2.2\pm0.5$ & 7.1$\times 10^{-7}$ & $2.5\pm0.7$  & 4.1$\times 10^{-1}$ & 3.8 \\
Ly$\gamma$ & 972.9  & $3.2\pm0.6$ & 3.9$\times 10^{-8}$ & $2.4\pm0.5$  & 1.1$\times 10^{-1}$ & 6.8 \\
\ion{C}{3} & 977.2  & $1.4\pm0.6$ & 7.6$\times 10^{-1}$ & $2.8\pm0.5$  & 7.2$\times 10^{-1}$ & 1.6 \\
\ion{N}{3} & 991.7  & $<1.5$      &  -                  & $0.8\pm1.1$  & 9.2$\times 10^{-1}$ & 0.8 \\
Ly$\beta$  & 1026.0 & $10.6\pm0.4$& $<10^{-10}$         & $8.9\pm0.4$  & 2.0$\times 10^{-6}$ & 12.7 \\
\ion{C}{2} & 1036.4 & $1.1\pm0.4$ & 2.8$\times 10^{-1}$ & $0.3\pm0.6$  & 1.0$\times 10^{0}$ & 1.5 \\
\ion{N}{2} & 1085.3 & $1.5\pm0.4$ & 3.2$\times 10^{-4}$ & $0.9\pm0.7$  & 8.0$\times 10^{-1}$ & 0.8 \\
\ion{C}{3} & 1175.8 & $1.3\pm0.4$ & 3.0$\times 10^{-3}$ & $2.2\pm0.7$  & 8.2$\times 10^{-2}$ & 0.8\\
\ion{Si}{2}& 1263.0 & $<1.0$      &  -                  & $<2.7$       & - & 1.4 \\
\ion{C}{2} & 1335.2 & $0.6\pm0.3$ & 5.5$\times 10^{-1}$ & $<2.5$       & - & 0.6 \\
\ion{Si}{4}& 1394.1 & $1.1\pm1.1$ & 8.7$\times 10^{-2}$ & $0.3\pm0.4$  & 0.99$\times 10^{-1}$ & 0.3 \\
\ion{Si}{4}& 1402.5 & $1.2\pm0.6$ & 6.5$\times 10^{-2}$ & $<2.0$       & - & 0.1 \\
\ion{C}{4}& 1550.2 & $<1.4$ & -  & $<2.7$       & - & 0.2 \\
\tableline 
\tablenotetext{a}{Uncertainties are from the 1$\sigma$ errors, and upper limits are for the 2$\sigma$ confidence level.}
\tablenotetext{b}{Best fit EHB + PAGB model for M~60, with $Z_{atm}$~=~0.1~$Z_{\sun}$.}
\end{tabular}
\end{table}

\end{document}